\documentclass[prb,preprint,preprintnumbers,amsmath,amssymb,unsortedaddress]{revtex4}

\usepackage{graphicx}
\usepackage{dcolumn}
\usepackage{bm}

\usepackage{color}

\begin{document}
 

\title{Current sheet formation and non-ideal behaviour at three-dimensional magnetic null points}

\author{D.~I.~Pontin} 
\email[]{dpontin@maths.dundee.ac.uk}
\altaffiliation[Now at: ]{Division of Mathematics, University of Dundee, Dundee, Scotland}
\author{A.~Bhattacharjee}
\affiliation{Space Science Center and Center for Magnetic Self-Organization, University of New Hampshire, Durham, New Hampshire, USA}
\author{K.~Galsgaard}
\affiliation{Niels Bohr Institute, University of Copenhagen, Copenhagen,
  Denmark}

\date{\today}

\begin{abstract}
The nature of the evolution of the magnetic field, and of current sheet
formation, at three-dimensional (3D) magnetic null points is investigated. A
kinematic example is presented which demonstrates that 
for certain evolutions of a 3D null (specifically those for which the ratios of the null point eigenvalues are time-dependent) there is no possible choice of boundary
conditions which renders the evolution of the field at the null ideal. 
Resistive MHD
simulations are described which demonstrate that such evolutions are generic.  A
3D null is subjected to boundary driving by shearing motions, and it is shown
that a current sheet localised at the null is formed. The qualitative and
quantitative properties of the current sheet are discussed. Accompanying the
sheet development is the growth of a localised parallel electric field,
one of the signatures of magnetic reconnection.
Finally, the relevance of the results to a recent theory of turbulent
reconnection is discussed.
\end{abstract}


\maketitle

\section{Introduction}
The three-dimensional topological structure of many astrophysical plasmas,
such as the solar corona,  is known to be highly complex. In order to
diagnose likely sites of energy release and dynamic phenomena in such
plasmas, where the magnetic Reynolds numbers are typically very large, it is
crucial to understand at which locations strong current concentrations will
form. These locations may be sites where singular currents are present
under an ideal magnetohydrodynamic (MHD) evolution.
In ideal MHD, the magnetic field is `frozen into' the plasma, and plasma
elements may move along field lines, but may not move across them. An
equivalent statement of ideal evolution is that
the magnetic flux through any material loop of plasma elements is conserved.

Three-dimensional (3D) null points and separators (magnetic field lines which
join two such nulls) 
might be sites of preferential current growth, in both the solar corona
\cite{longcope1996,priest2005} and the
Earth's magnetosphere, see Refs.~[\onlinecite{priest2000,siscoe1988}] for reviews. 3D
null points are thought to be present in abundance 
in the solar corona. A myriad of magnetic flux concentrations penetrate the
solar surface, and it is predicted that for every 100 photospheric
flux concentrations, there should be present between approximately 7 and 15
coronal null points \cite{schrijver2002,longcope2003,close2004}. 
Furthermore, there is observational evidence that reconnection involving a 3D
null point  may be at work in  some solar flares \citep{fletcher2001} and
solar eruptions \cite{aulanier2000,ugarteurra2007}. Closer to home, there has
been a recent in situ observation \citep{xiao2006} by the Cluster spacecraft
of a 3D magnetic null, which is proposed to play an important role in the
observed signatures of reconnection within the Earth's magnetotail. In addition,
current growth at 3D nulls has been observed in the laboratory
\cite{bogdanov1994}. 

While the relationship between reconnection at a separator (defined by a
null-null line) and reconnection at a single 3D
null point is not well understood, it is clear that the two should be linked in
some way. Kinematic models anticipate that null points and separators are both
locations where current singularities may form in ideal MHD
\cite{lau1990,priest1996}. Moreover, it is also expected that 3D nulls
\cite{bulanov1984,pontincraig2005} and  separators \cite{longcopecowley1996}
may each collapse to singularity in response to external motions. 

The linear field topology in the vicinity of a 3D magnetic null point may be
examined by considering a Taylor expansion of ${\bf B}$ about the null;
\begin{displaymath}
{\bf B} = {\cal M} \cdot {\bf r},
\end{displaymath}
where the matrix ${\cal M}$ is the Jacobian of ${\bf B}$ evaluated at the
null \cite{fukao1975,parnell1996}. The eigenvalues of ${\cal
  M}$ sum to zero since $\nabla \cdot {\bf B}=0$. The two eigenvectors
corresponding to the eigenvalues with
like-sign real parts define the fan (or $\Sigma$-) surface of the null, in
which field lines approach (recede from) the null. The third eigenvector
defines the orientation of the spine (or $\gamma$-) line, along which a single
pair of field lines recede from (approach) the null [see Fig.~\ref{nullpic}(a)].

In Section \ref{kinematic} we discuss the nature of the evolution of the
magnetic 
field in the vicinity of a 3D null, and provide an example which demonstrates
that certain evolutions of the null are prohibited in ideal MHD. In Sections
\ref{numsec} and \ref{quantsec} we present results of numerical simulations of
a 3D null which is 
driven from the boundaries, and describe the qualitative and quantitative
properties of the resulting current sheet. In Section \ref{turbsec} we observe
that our models may point towards 3D nulls as a possible site where turbluent
reconnection might take place. Finally in Section \ref{conc} we give a
summary.


\section{Non-ideal evolution around 3D nulls}\label{kinematic} 
\subsection{Ideal and non-ideal evolution}\label{boozer} 
The evolution of a magnetic field is said to be
ideal if ${\bf B}$ can be viewed as being frozen into some ideal flow, i.e.~there
exists some ${\bf w}$ which satisfies
\begin{equation}\label{idealohm}
{\bf E} + {\bf w} \times {\bf B} = -\nabla \Phi,
\end{equation}
or equivalently
\begin{equation}\label{idealind}
\frac{\partial {\bf B}}{\partial t}=\nabla \times \left( {\bf w} \times {\bf 
    B} \right) 
\end{equation}
everywhere. In order for the evolution to be ideal, ${\bf w}$ should be continuous and
smooth, such that it is equivalent to a real plasma flow. Examining the
component of Eq.~(\ref{idealohm}) parallel to ${\bf B}$, it is clear that in
configurations containing closed field lines, the constraint that $\oint {\bf
  E}\cdot {\bf dl}=0$ must be satisfied in order to satisfy
Eq.~(\ref{idealohm}), so that $\Phi$ is  single-valued. However, even when no
closed field 
lines are present, there are still configurations in which it may not be
possible to find a smooth velocity ${\bf w}$ satisfying Eq.~(\ref{idealohm}). 
These configurations may contain isolated null points, or pairs of null points
connected by separators \cite{lau1990, priest1996, longcopecowley1996}.  We
will concentrate in what follows on the case where isolated null points are
present.  

Even when non-ideal terms are present in Ohm's law, it may still not be
possible to find a smooth unique velocity ${\bf w}$ satisfying
Eq.~(\ref{idealind}) if the non-ideal region is embedded in an ideal region
(i.e.~is spatially localised in all three dimensions), as this imposes a
`boundary condition' on $\Phi$ within the (non-ideal) domain. If field lines
link different points (say ${\bf x}_1$ and ${\bf x}_2$) on the boundary
between the ideal and non-ideal region, then this imposes a constraint on
allowable solutions for $\Phi$ within the non-ideal region, since they must be
consistent with $\Phi({\bf x}_1)=\Phi({\bf x}_2)$
\cite{hornig2003,priesthornig2003}.

It has recently been  claimed by Boozer \cite{boozer2002} that the evolution of the
magnetic field in the vicinity of an isolated 3D null point can always be
viewed as ideal. However, we will demonstrate in the following section that in
fact certain evolutions of a 3D null are prohibited under ideal
MHD, and must therefore be facilitated by non-ideal processes. 
Furthermore, in subsequent sections we will show that such
evolutions are a natural consequence of typical perturbations of a 3D null. 

The argument has been proposed \citep{boozer2002} that for generic 3D nulls
\begin{equation}\label{wrongeq}
\nabla \times \left( {\bf w}^{\prime} \times {\bf B} \right) \equiv 
{\bf w}^{\prime} \cdot \nabla {\bf B} = 
- \left(\frac{\partial {\bf B}}{\partial t}\right)_{{\bf x}_0} 
\end{equation}
at the null point, where ${\bf x}_0$ is the position of the null, which can
  always be solved for ${\bf w}^{\prime}$ so long 
  as the matrix $\nabla {\bf B}$ is invertible. For a `generic' 3D null with
  ${\rm det}(\nabla {\bf B}) \neq 0$ this is always possible. However,
  Eq.~(\ref{wrongeq}) is only valid {\it at the null itself}, with 
solution ${\bf w}^{\prime}=d{\bf x}_0 / dt$. In general, $\nabla \times
\left( {\bf w}^{\prime} \times {\bf B} \right) \neq {\bf w}^{\prime} \cdot
  \nabla {\bf B}$ {\it 
    in the vicinity} of a given point. That is, it is not valid to
  discard the 
terms ${\bf B} (\nabla \cdot {\bf w}^{\prime})$ and $({\bf B}\cdot\nabla) {\bf
  w}^{\prime}$ from the right-hand side of Eq.~(\ref{idealind}) (when expanded
  using the appropriate identity) when considering the behaviour of
  the flow around the null. In order to describe the evolution of
  the field around the null, we must consider the evolution in some finite
  (possibly infinitesimal) volume about the null.   
Under the assumption (\ref{wrongeq}), the vicinity of the null moves like a
`solid body'---which seems to preclude any field evolution. Thus, the
equation simply states that the field null remains, supposing no bifurcations
occur---it shows that the null point cannot disappear, but does not describe
the flux velocity in any finite volume around the null. This argument---that
  the velocity of the null and the value of the flux-transporting velocity at
  the null need not necessarily be the same---has been
  made regarding 2D X-points by Greene \cite{greene1993}.

Furthermore, in order to show that the velocity ${\bf w}^{\prime}$ is non-singular
at the null, the assumption  is made in Ref.~[\onlinecite{boozer2002}] that $\Phi$ can
be uniquely defined 
at the null, and that it may be approximated by a Taylor expansion
about the null, which clearly presumes that $\Phi$ is a well-behaved function. 
As will be shown in what follows, in certain situations, it is
not possible to find a function $\Phi$ which is (or in particular whose gradient
$\nabla \Phi$ is) well-behaved at the null point (smooth and continuous). This
is demonstrated below by analysing the properties of the function $\Phi$ with
respect to  {\it any} arbitrary boundary conditions 
away from the null point itself. By beginning at the null point and
extrapolating $\Phi$ outwards, it may indeed be possible to choose $\Phi$ to be
well-behaved at the null, but this $\Phi$ must not be consistent with
any physical (i.e.~smooth) boundary conditions on physical quantities such as
the plasma flow.

In fact, it has been proven by Hornig and Schindler \cite{hornig1996} that no
smooth `flux velocity' or velocity of `ideal evolution' exists satisfying
Eq.~(\ref{idealind}) if the ratios of the null point eigenvalues change in
time. They have shown that under an ideal evolution, the 
eigenvectors (${\bf q}_{\alpha}$)  and eigenvalues ($\alpha$) of the null
evolve according to
\begin{equation}\label{eigenevol}
\frac{d {\bf q}_{\alpha}}{dt}= {\bf q}_{\alpha} \cdot \nabla {\bf w},  \qquad
\frac{d \alpha}{dt} = C \alpha,
\end{equation}
where $C=-\nabla\cdot {\bf w} |_{{\bf x}_0}$ is a constant. From the second
equation it is straightforward to show that the ratio of any two of the eigenvalues
must be constant in time.

Note finally that the assumption that ${\rm det}(\nabla {\bf B}) \neq 0$ rules
out the possibility of bifurcation of null points. While
it is non-generic for ${\rm det}(\nabla {\bf B}) = 0$ to persist for a finite
period of 
time, when multiple null points are created or annihilated in a bifurcation
process, a higher order null point is present right 
at the point of bifurcation, and ${\rm det}(\nabla {\bf B})$ passes
through zero. Such bifurcation processes are naturally occurring, and clearly
require a reconnection-like process to take place.


\subsection{Example}\label{kinex}
We can gain significant insight by considering the kinematic problem. We
consider an ideal situation, so that Ohm's law takes the form of
Eq.~(\ref{idealohm}). Uncurling Faraday's law gives ${\bf E} = -\nabla \Phi^{\prime} -
\frac{\partial {\bf A}}{\partial t}$, where ${\bf B}=\nabla\times{\bf A}$, and
combining the two equations we have that
\begin{equation}\label{master}
\nabla \widetilde{\Phi} + {\bf w} \times {\bf B} = \frac{\partial {\bf A}}{\partial t},
\end{equation}
where $\widetilde{\Phi}=\Phi-\Phi^{\prime}$. We proceed as follows. We consider a given
time-dependent magnetic field, and calculate the corresponding functions
$\widetilde{\Phi}$ and ${\bf w}_{\perp}$. The component of ${\bf w}$ parallel to ${\bf B}$
is arbitrary with respect to the evolution of the magnetic flux.
We choose the  magnetic field such that we have an analytical parameterisation
of the field lines,
obtained by solving $\frac{d {\bf X}}{ds} = {\bf B}({\bf X}(s))$, where the
parameter $s$ runs along the field lines. Then taking the component of
Eq.~(\ref{master}) parallel to ${\bf B}$ we have
\begin{equation}
\begin{array}{rcl}
\nabla \widetilde{\Phi} \cdot {\bf B} & = & \frac{\partial {\bf A}}{\partial t} \cdot {\bf B}\\ 
\widetilde{\Phi} & = & \int \frac{\partial {\bf A}}{\partial t} \cdot {\bf B} \: ds,
\end{array}
\end{equation}
where the spatial distance $dl=|{\bf B}|ds$. Expressing
${\bf B}$ and $\frac{\partial {\bf A}}{\partial t}$ as functions of $({\bf
  X}_0,s)$ allows us to perform the integration. This defines
$\nabla_{\parallel} \widetilde{\Phi}$, 
while the constant of this integration, $\widetilde{\Phi}_0({\bf X}_0)$, allows for different
$\nabla_{\perp}\widetilde{\Phi}$. We now
substitute in the inverse of the field line equations 
${\bf X}_0({\bf X})$ to find $\widetilde{\Phi}({\bf X})$. Finally,
\begin{equation}\label{vperpeq}
{\bf w}_{\perp} = -\frac{(\frac{\partial {\bf A}}{\partial t} - \nabla
\widetilde{\Phi}) \times {\bf B}}{B^2}.
\end{equation}

Consider now a null point, the ratios of whose eigenvalues change in time. 
Take the magnetic field to be
\begin{equation}
{\bf B} = \left( -2x - y f(t), y+ x f(t), z \right).
\end{equation}
A suitable choice for $\frac{\partial {\bf A}}{\partial t}$ is $\frac{\partial
  {\bf A}}{\partial t}= -f^{\prime}/2 \: (0, 0, x^2+y^2)$ 
(other appropriate choices lead to the same conclusions as below). 
The eigenvalues of  the null point are 
\begin{equation}
\lambda_1=1 \qquad,\qquad \lambda_{\pm} = \frac{-1 \pm \sqrt{9-4f^2}}{2},
\end{equation}
with corresponding eigenvectors
\begin{eqnarray}
{\bf q}_1   & = & (0,0,1) \nonumber\\
{\bf q}_{+} & = & \left(\frac{\lambda_{+}-1}{f},1,0 \right) \nonumber\\
{\bf q}_{-} & = & \left(1,\frac{f}{\lambda_{-}-1},0 \right). \nonumber
\end{eqnarray}

From the above it is clear that the spine (defined by $(1,0,0)$
when $f=0$) is given by
\begin{displaymath}
z=0 \quad , \quad (1-\lambda_{-})y + fx =0.
\end{displaymath}
In addition, the equation of the fan plane (which lies at $x=0$ when $f=0$)
can be found by solving ${\hat {\bf n}} \cdot {\bf r}=0$, $\hat{{\bf n}}= {\bf q}_1 \times {\bf q}_{+}$ to give
\begin{displaymath}
\frac{(1-\lambda_{+})}{f} \: y + x = 0.
\end{displaymath}
Thus the spine and fan of the null close up on one another in time (being
perpendicular when $f(t)=0$), see Fig.~\ref{wnonsmooth}(b). Note that we require $|f|<3/2$ to preserve the
nature of the null point, 
otherwise is collapses to a non-generic null; we may take for example $f={\rm
  tanh}(t)$. 
In order to simplify what follows, we define two new functions $M$ and $P$
which describe the locations of the spine and fan of the null:
\begin{eqnarray}
M(x,y,t) & = & \frac{(1-\lambda_{-})y + fx}{f x_1},\\
P(x,y,t) & = & \frac{(1-\lambda_{+})y + fx}{f x_1}.
\end{eqnarray}

\subsubsection{Analytical field line equations}
Now, the first step in solving the equations is to find a representation of
the field lines. Solving $\frac{d {\bf X}}{ds} = {\bf B}({\bf X}(s))$ gives
\begin{eqnarray}
x & = & (\lambda_{+}-1) C_1 e^{\lambda_{+} s}  +  (\lambda_{-}-1) C_2 
e^{\lambda_{-} s}.\label{zz} \nonumber\\
y & = & f C_1 e^{\lambda_{+} s}  +  f C_2 e^{\lambda_{-} s} \label{yy} \label{Bl}\\
z & = & z_0 e^s \label{xx} \nonumber
\end{eqnarray}
where $C_1$ and $C_2$ are constants. Clearly the equation for $z$ is simple,
but the other two are coupled in a more complicated way. 
In order to render the field line equations invertible, we choose to
set $s=0$ on surfaces which move in time, tilting as the null point
does. This is allowable since there is no linking between the integration of
the field lines and the time derivative, that is, $t$ (or $f$) is just a
constant in the integration. We choose to set $s=0$ on surfaces defined by
\begin{equation}\label{tiltsurf}
x = x_1 - \frac{1 - \lambda_{+}}{f} y \quad , \quad y=y_0 - \frac{x_1 f}{\lambda_{+} -
  \lambda_{-}}
\end{equation}
where $x_1$ is a constant and $y_0=y_0(x,y,t)$ is the starting position of the
field line footpoint on the tilting surface. Then
\begin{equation}\label{Cs}
C_1 = \frac{y_0}{f} \quad , \quad C_2 = \frac{-x_1}{\lambda_{+} - \lambda_{-}}.
\end{equation}
We can see by comparison of equations (\ref{tiltsurf}) and (\ref{Cs}) that
our `boundaries' lie parallel to the fan plane, but with a shift of
$\pm x_1$.

With $C_1$ and $C_2$ now defined by Eq.~(\ref{Cs}), Eqs.~(\ref{Bl}) may be
inverted to give 
\begin{eqnarray}
y_0 & = & \frac{fx_1}{\lambda_{+}-\lambda_{-}} M 
P^{\lambda_{+}/\lambda_{-}}, \label{inv_Bl}\nonumber\\
z_0 & = & z P^{-1/\lambda_{-}}, \\
s & = & \frac{1}{\lambda_{-}} {\rm ln} P. \nonumber 
\end{eqnarray}


\subsubsection{Solving for $\widetilde{\Phi}$ and physical quantities}
The first step in determining the solution is to solve for
$\widetilde{\Phi}$, which is given by
\begin{equation}
\widetilde{\Phi} = \int_0^s z (x^2 + y^2) \: ds
\end{equation}
where $x$, $y$, $z$ are functions of $(y_0, z_0, s)$. Upon substitution
of (\ref{Bl}), it is 
straightforward to carry out the integration in a symbolic computation package
(such as Maple or Mathematica), since the integrand is simply a
combination of exponentials in $s$. We then substitute (\ref{inv_Bl}) into the
result to obtain $\widetilde{\Phi}(x,y,z)$.

\begin{figure}
\centering
\includegraphics{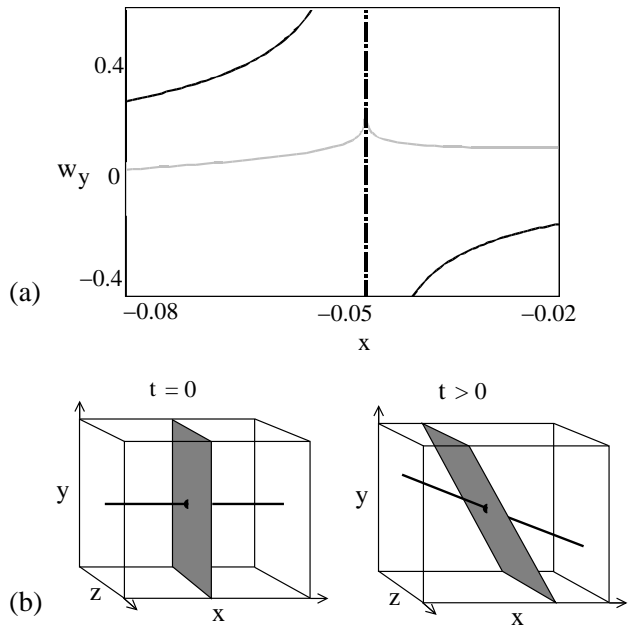}
\caption{(a) Plots of $w_y$ against $x$ for $\widetilde{\Phi}_0=0$ (black line) and for the case where
  $\widetilde{\Phi}_0$ is defined by Eq.~(\ref{phi0}) (grey line). The dashed line
  indicates the location of the fan plane, and we take $f=0.2, x_1=1, y=0.7, z=0.5$. (b) Orientation of the null spine and fan for $t=0$ (left) and $t>0$ (right).}
\label{wnonsmooth}
\end{figure}
For any general choice of the function $\widetilde{\Phi}_0$ (i.e.~choice of initial
conditions for the integration, which may be viewed as playing the role of
boundary conditions), we find that $\widetilde{\Phi}$ is non-smooth at
the fan plane, and therefore $\nabla \widetilde{\Phi}$ tends to infinity there. Thus the
electric field ${\bf E}$ and flux velocity ${\bf w}$ will tend to infinity at the
fan plane (see Fig.~\ref{wnonsmooth}(a)). Examining the
expression for $\widetilde{\Phi}$,  it is apparent that
the problematic terms are those in $P^{-1/\lambda_{-}}$ (since $-1/\lambda_{-}
<-1/2$ for $-1\leq f \leq 1$). In order to cancel out all of these terms, we must
take (by inspection)
\begin{equation}\label{phi0}
\widetilde{\Phi}_0 = -x_1^2 \: \frac{f^2 + (1-\lambda_{-})^2}
{(\lambda_{+}-\lambda_{-})^2 (1+2 \lambda_{-}) } \: z_0.
\end{equation}
With this choice, $\widetilde{\Phi}$ is smooth and continuous everywhere. However, $\nabla
\widetilde{\Phi}$ is still non-smooth in the fan plane. Again examining $\widetilde{\Phi}$, we see that this
results from terms in $P^{\mu_1} {\rm ln} P$ ($\mu_1$ constant). These take the form (after
simplification) 
\begin{eqnarray}
\widetilde{\Phi} &=& ......-4fx_1 y_0 z_0 s \nonumber\\
     &=& ......-\frac{4f^2 x_1^2}{\lambda_{+}-\lambda_{-}} \: z
P^{-1/\lambda_{-}} \: 
M P^{-\lambda_{+}/\lambda_{-}} \: {\rm ln}P^{-1/\lambda_{-}}.
\end{eqnarray}
Now, due to the form of $y_0(x,y,z,t)$ and $z_0(x,y,z,t)$ (Eq.~\ref{inv_Bl}),
it is impossible to remove the term in $P^{\mu_1} {\rm ln} P$ through addition of
any $\widetilde{\Phi}_0 (y_0, z_0)$ without inserting some term in either $z^{\mu_2} {\rm
  ln}z$ or $M^{\mu_3} {\rm ln}M$ ($\mu_i$ constant). This is
simply equivalent to
transferring the non-smoothness to the vicinity of the spine instead of the
fan. Thus, there is no choice of boundary conditions which can
possibly render the evolution `ideal'. 

The example discussed above demonstrates that in an ideal plasma, it is
not possible for a 3D null point to evolve in the way described, with the
spine and fan opening/closing towards one another. Therefore, if
such an evolution is to occur, non-ideal processes must become
important.


\section{Resistive MHD simulations}\label{numsec}

We now perform numerical simulations in the 3D resistive MHD model. 
The setup of the simulations is very similar to that described by Pontin and Galsgaard
\cite{pontingalsgaard2007}. More details of the numerical scheme may be found in 
Refs.~[\onlinecite{galsgaard1997,archontis2004}]. 
We consider an isolated 3D null point within our computational volume, which is
driven from the boundary. We focus on the case where the null
point is driven from its spine footpoints. The driving takes the form of a
shear. We begin initially with a potential null point; ${\bf B}=B_0(-2x, y,
z)$, and with the density $\rho=1$, and the internal energy
$e=5\beta/2$ within the domain, so that we initially have an equilibrium. Here
$\beta$ is a constant which determines the plasma-$\beta$, which is infinite
at the null itself, but decreases away from it.
We assume an ideal  gas, with $\gamma=5/3$, and take $B_0=1$ and $\beta=0.05$ in
each of the simulations. A stretched numerical grid is used to give higher resolution near the null (origin). 
Time units in the simulations are equivalent to the Alfv{\' e}n travel time across a unit length
in a plasma of density $\rho=1$ and uniform magnetic field of modulus 1. The resistivity is taken to be 
uniform, with its value being based upon the dimensions of the domain. Note that at $t=0$, 
${\bf B}$ is scale-free as it is linear, and thus, the actual value of $\eta$ is somewhat arbitrary until we fix a physical length scale to associate with the size of our domain.
\begin{figure}
\centering
(a)\includegraphics[scale=0.37]{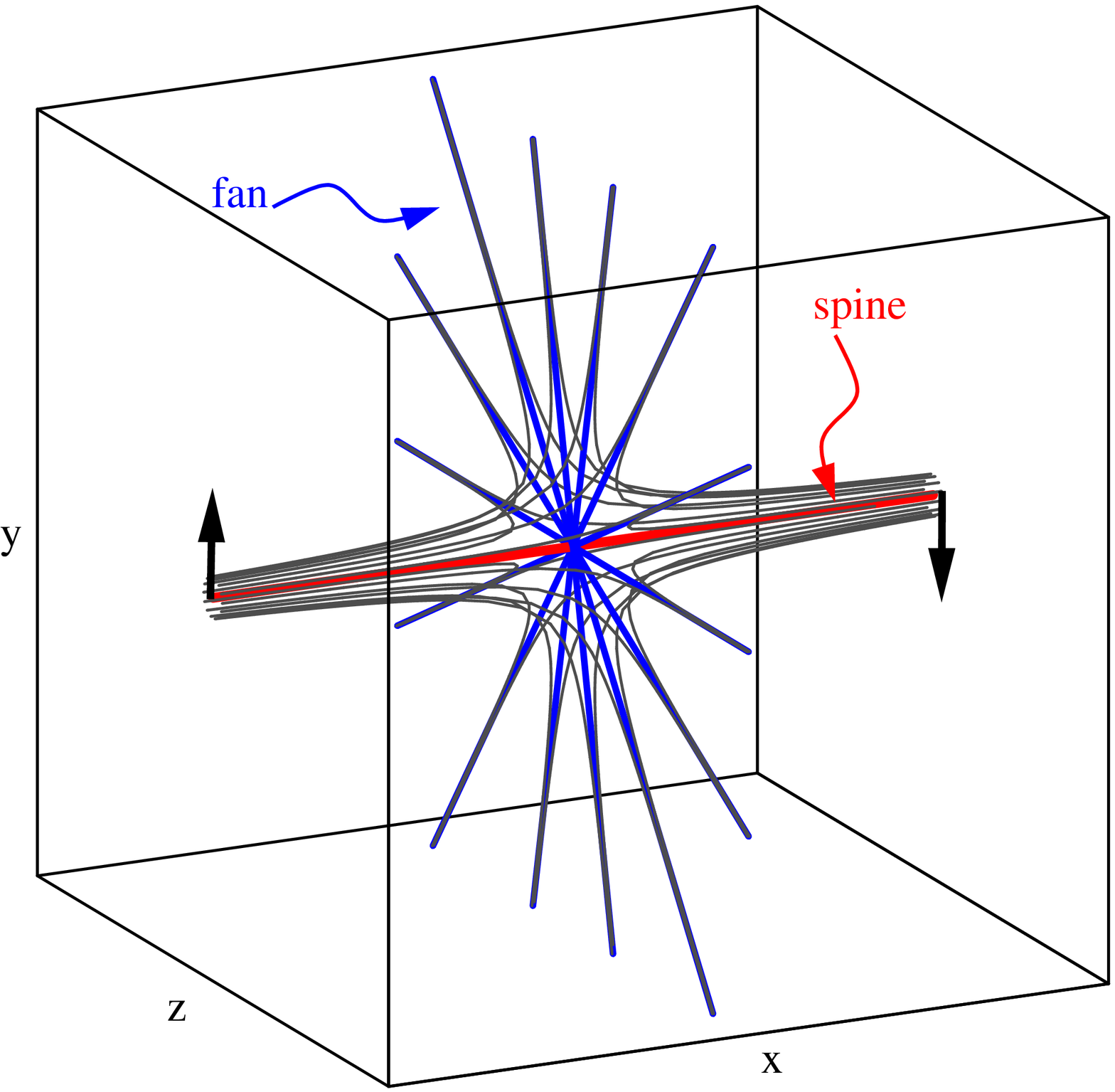}
(b)\includegraphics[scale=0.6]{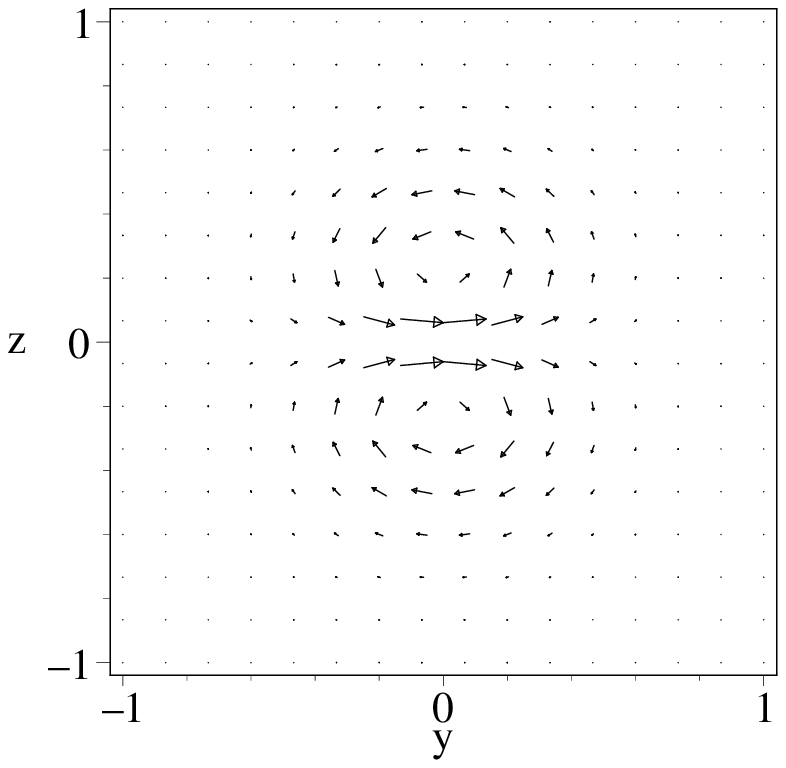}
\caption{(a) (Colour online) Schematic of the 3D null point in the computational domain. The black
arrows indicate the direction of the boundary driving. (b) Boundary driving
flow at $x=-X_L$, the lower $x$-boundary, for $Y_l=Z_l=3$ and $A_d=80$.}
\label{nullpic}
\end{figure}

At $t=0$, the
spine of the null point is coincident with the $x$-axis, and the fan plane
with the $x=0$ plane [see Fig.~\ref{nullpic}(a)].
A driving velocity is then imposed on the (line-tied) $x$-boundaries, which
advects the spine footpoints in opposite directions  
on the opposite boundaries (chosen to be in the $y$-direction without loss
of generality). 
We choose an incompressible (divergence-free) velocity field, defined by the stream
function 
\begin{equation}\label{incompprof}
\psi = V_0(t) \cos^2\left( \frac{\pi y_1}{2} \right) 
\sin \left( \pi z_1 \right) e^{-A_d (y_1^2+z_1^2)},
\end{equation}
where $y_1=y/Y_l$ and $z_1=z/Z_l$, the numerical domain has dimensions $[\pm
X_l, \pm Y_l, \pm Z_l]$, and $A_d$ describes the 
localisation of the driving patch. This drives the spine footpoints in the
$\pm {\hat {\bf y}}$ direction, but has return flows at larger radius (see
Fig.~\ref{nullpic}(b)). Note however that the field lines in the 
return flow regions never pass close to the fan (only field lines
very close to the spine do). We have checked that there are no major
differences for the evolution from the case of uni-directional driving (${\bf
  v}=v_y{\hat{\bf y}}$). 

Two types of temporal variation for the driving are considered. In the first,
the spine is driven until the resulting disturbance reaches the null, forming
a current sheet (see below), then the driving switches off. It is smoothly
ramped up and down to reduce sharp wavefront generation. In the other case, the
driving is ramped up to a constant value and then held there (for as long as
numerical artefacts allow). Specifically, we take either
\begin{equation}\label{transientprof}
V_0(t) = v_0 \left( \left(\frac{t-\tau}{\tau}\right) ^4 -1 \right) ^2 \qquad,
\qquad 0 \leq t \leq \tau,
\end{equation}
or
\begin{equation}
V_0(t) = v_0 \tanh ^2 ( t/ \tau ),
\end{equation}
where $v_0$ and $\tau$ are constant. We begin by discussing the results of
runs with the transient driving profile, as described by
Eqs.~(\ref{incompprof}) and (\ref{transientprof}).


\subsection{Current evolution and plasma flow}\label{curevsec}

Unless otherwise stated, the following sections describe results of a
simulation run with the transient driving time-dependence, and with parameters
$X_l=0.5$, $Y_l=Z_l=3$, $A_d=80$, $v_0=0.01$, $\tau=1.8$ and $\eta=10^{-4}$. The resolution is $128^3$, and the grid spacing at the null is $\delta x \sim 0.005$ and $\delta y, \delta z \sim0.025$ (the results have been checked at $256^3$ resolution).

\begin{figure}
\centering
(a)\includegraphics[scale=1.1]{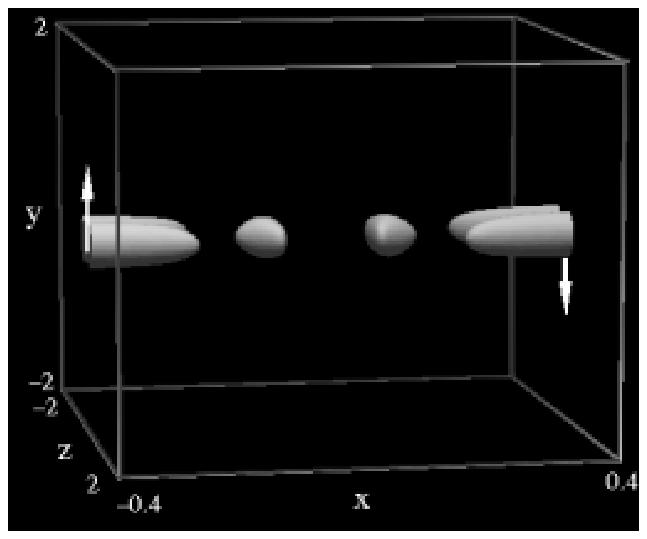}
(b)\includegraphics[scale=0.45]{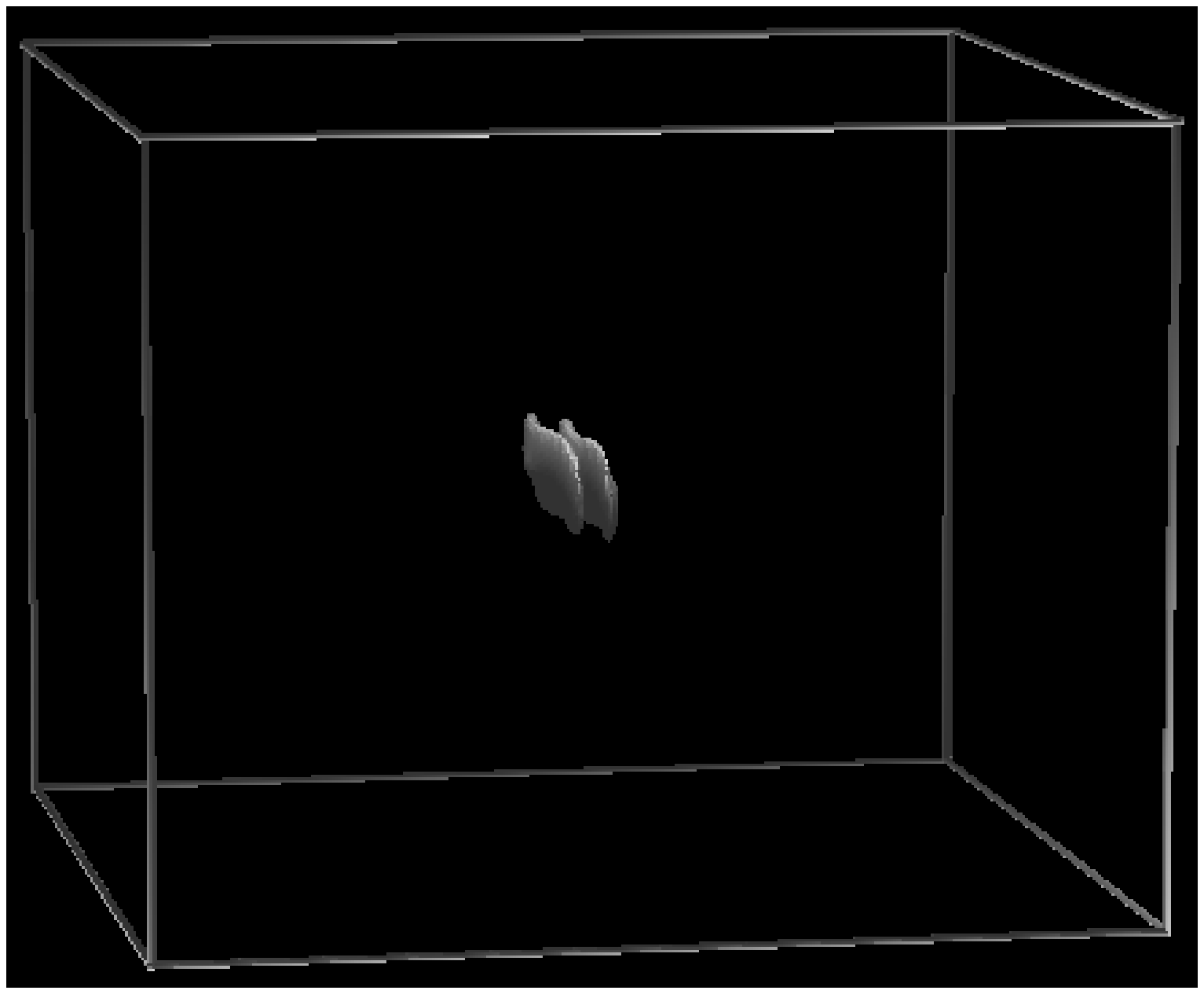}
(c)\includegraphics[scale=0.45]{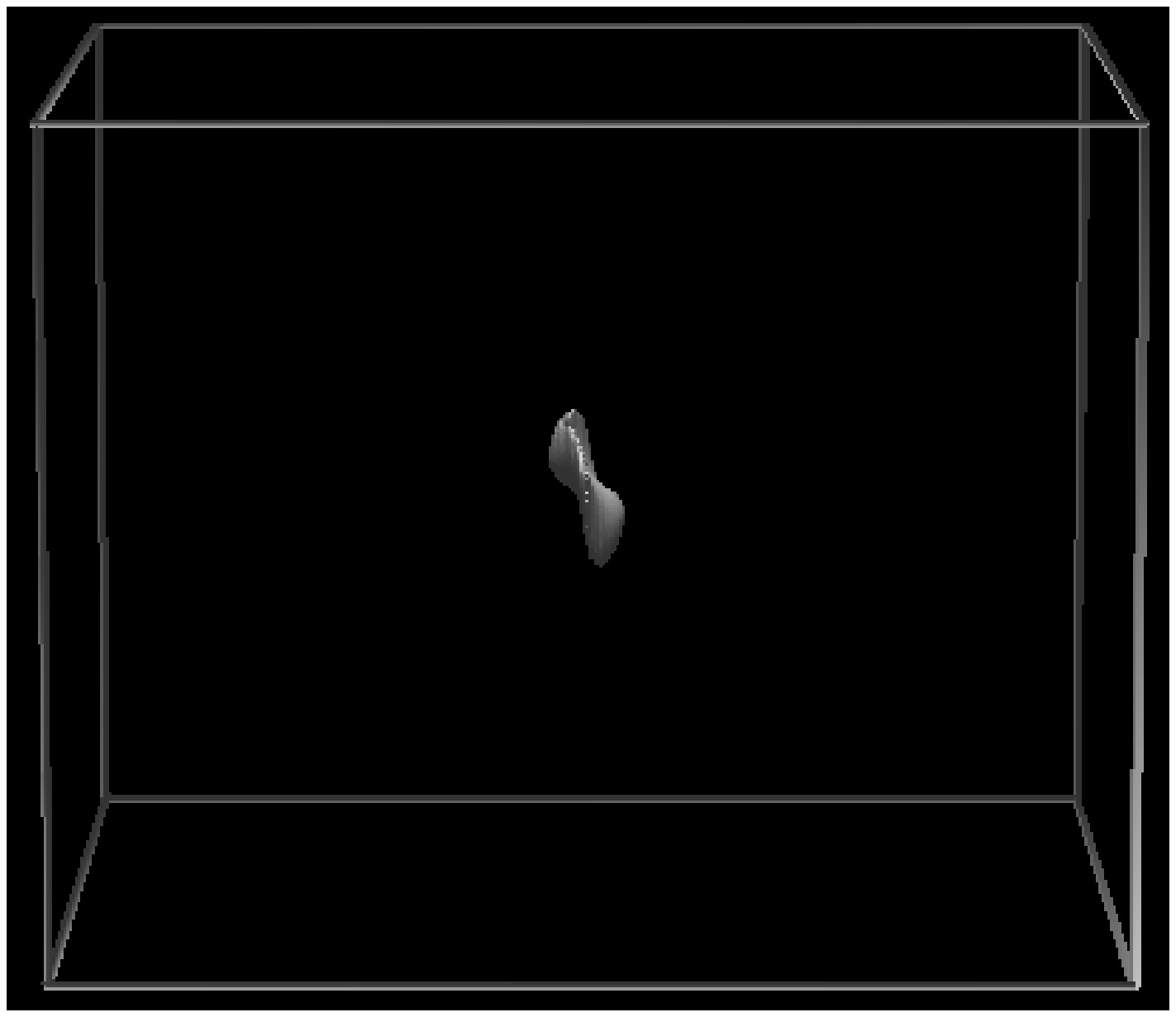}
(d)\includegraphics[scale=0.45]{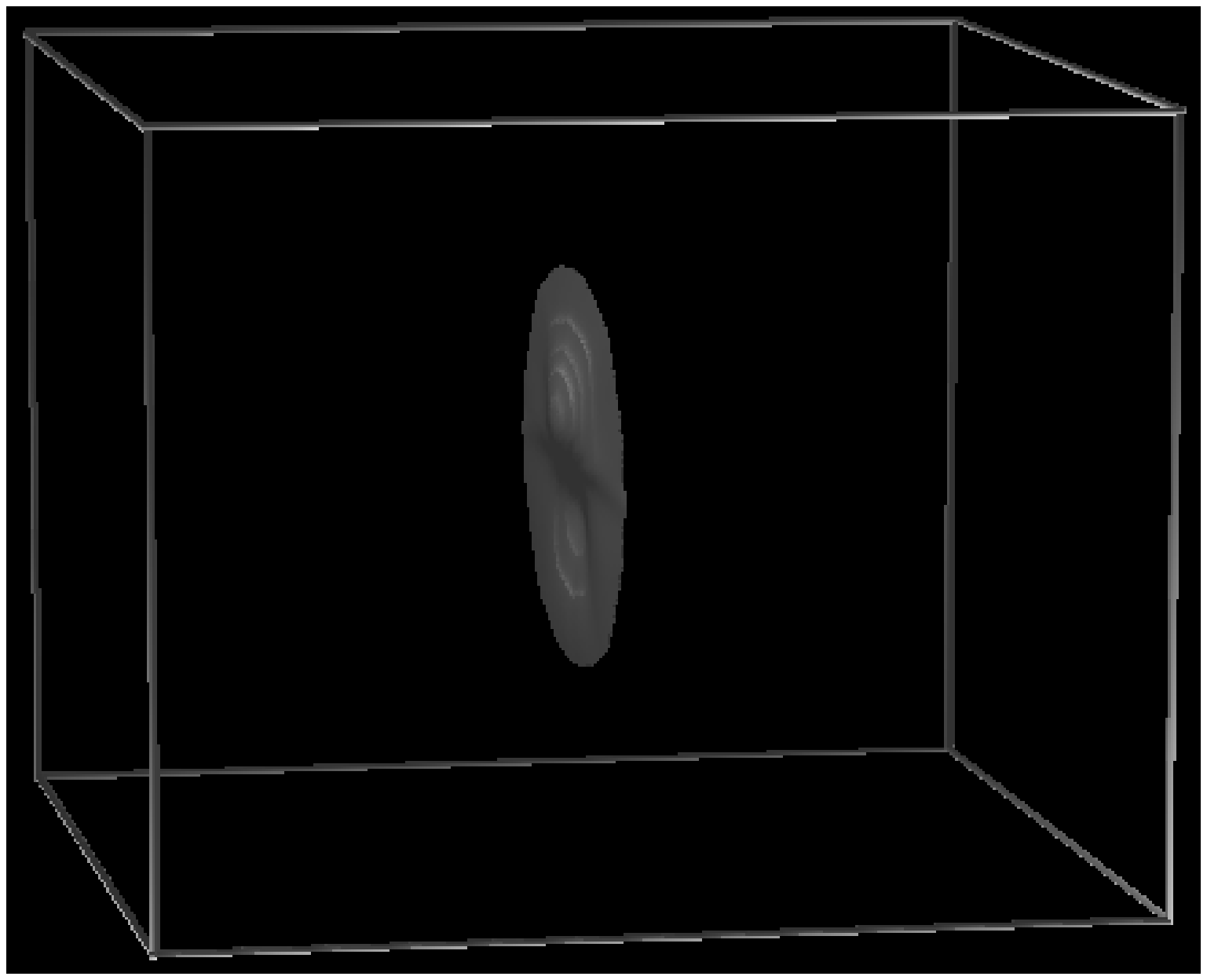}
\caption{Isosurfaces of $J$, at 50$\%$ of the maximum at that time, for times
marked by diamonds in Fig.~\ref{jcomptime} ($t=1,2,3,5$)}
\label{jisotrans}
\end{figure}

As the boundary driving begins, current naturally develops near
the driving regions. This disturbance propagates inwards towards the
null at the local value of the Alfv{\' e}n speed. 
It concentrates at the null point itself, with the maximum of $J=|{\bf J}|$
growing sharply once the disturbance has reached the null, where the Alfv{\'
  e}n velocity vanishes. Figure
\ref{jisotrans} shows isosurfaces of $J$ (at 50$\%$ of maximum).

\begin{figure}
\centering
\includegraphics[scale=0.858]{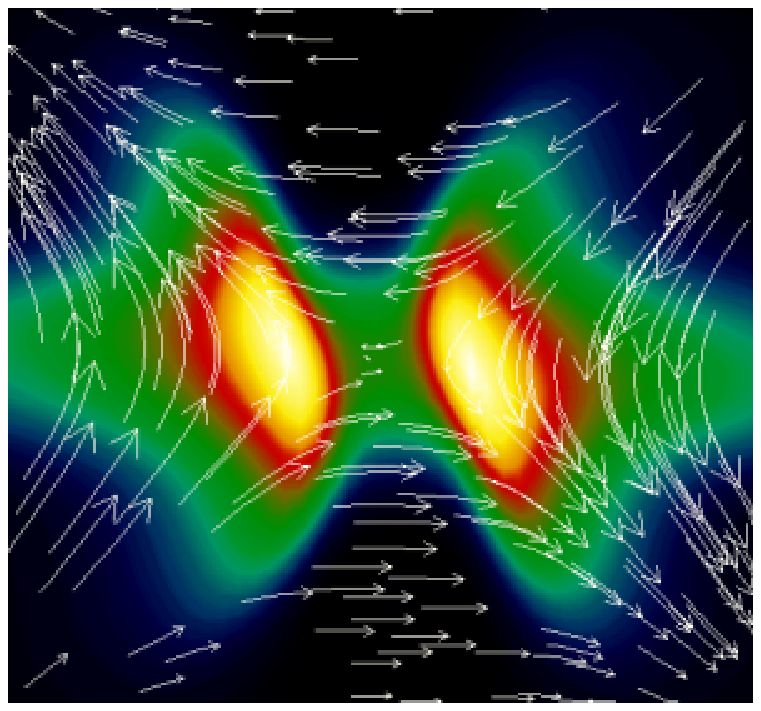}
\includegraphics[scale=1]{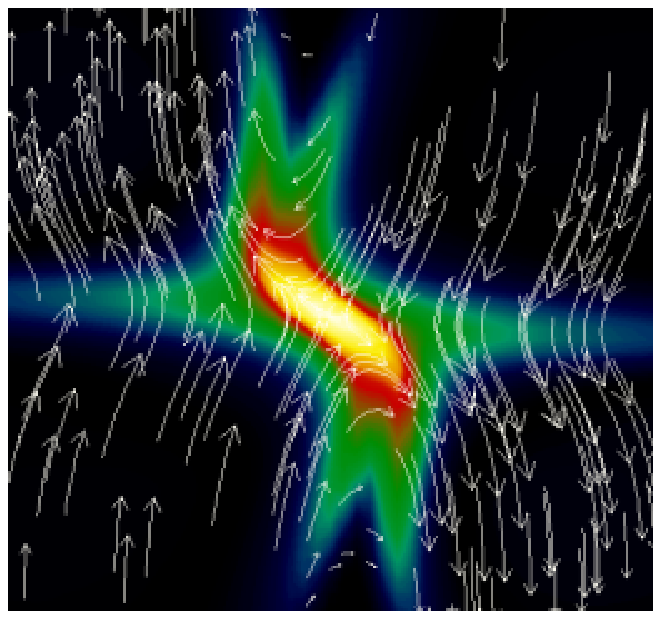}
\includegraphics[scale=1]{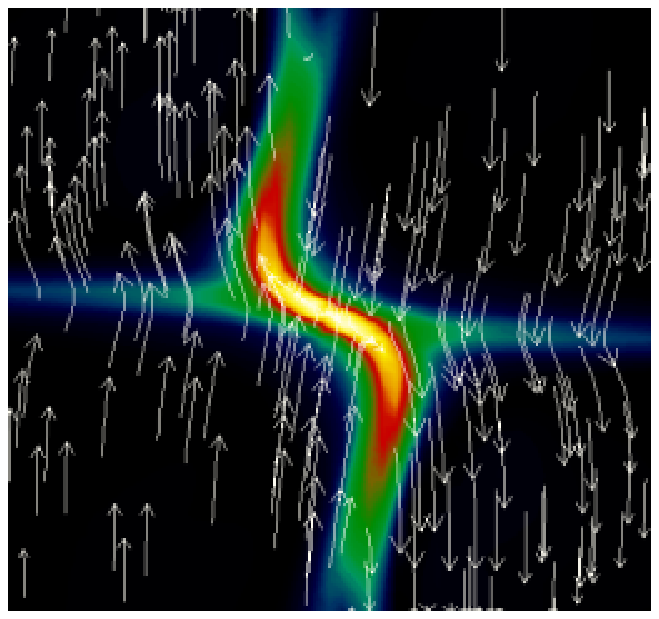}
\includegraphics[scale=1]{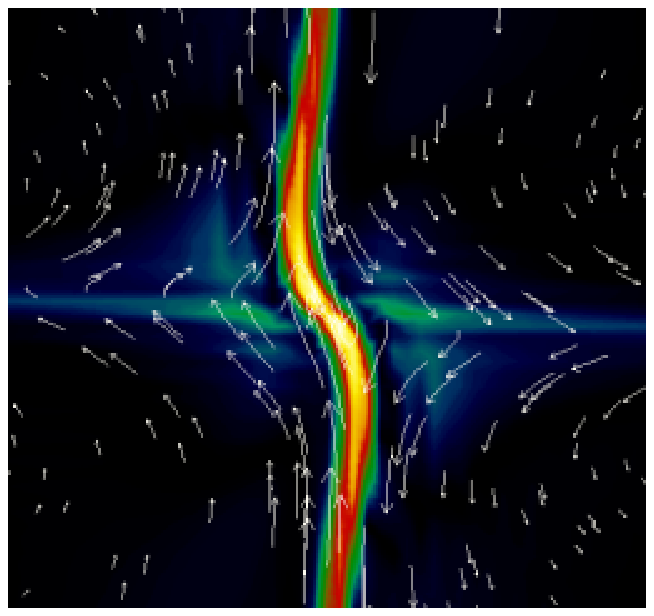}
\caption{(Colour online) Arrows show plasma flow, while shading shows $|J|$ (scaled to the
  maximum in each individual frame). Viewed in
  the $z=0$ plane, inner 1/4 of domain: $[x,y]=[-0.12..0.12, -0.7..0.7]$ (y
  vertical). Images are at $t=1.6,2.4,3.0,5.0$.} 
\label{vtransfig}
\end{figure}

It is fruitful to examine the current evolution in a plane of constant $z$ (plane of
the shear). In such a plane, the current forms
an `S'-shape, and is suggestive of a collapse from an X-type
configuration (spine and fan orthogonal) to a Y-type
configuration in the shear plane (Fig.~\ref{vtransfig}). Note that once the
driving switches off, the current begins to weaken 
again and spreads out in the fan plane, and the spine and fan relax back towards
their initial perpendicular configuration (see Figs.~\ref{jisotrans}(d),
\ref{vtransfig}(d)). 

That we find a current concentration forming which is aligned not to the
(global directions of the) fan or spine of the null (in contrast with simplified
analytical self-similar \cite{bulanov1984} and incompressible
\cite{craig1996} solutions), but at some intermediate 
orientation, is entirely consistent with the laboratory observations of
Bogdanov {\it et al.} \cite{bogdanov1994}. In fact the angle that our sheet
makes in the $z=0$ plane (with, say, the $y$-axis) is dependent on the
strength of the driving (strength of the current), as well as the field
structure of the `background' null (as found in Ref.~[\onlinecite{bogdanov1994}]) and
the plasma parameters. We
should point out that we consider the case where in the notation of Bogdanov
{\it et al.} 
$\gamma<1$. They consider a magnetic field ${\bf B}=\left((h+h_r)x,
-(h-h_r)y, -2h_r z\right)$, and define $\gamma=h_r/h$.
Thus, for $\gamma>1$ they have current parallel to the spine, which we
expect to be resultant from rotational motions, and have very different
current sheet and flow structures \cite{pontingalsgaard2007,pontin2004}.

\begin{figure}
\centering
\includegraphics{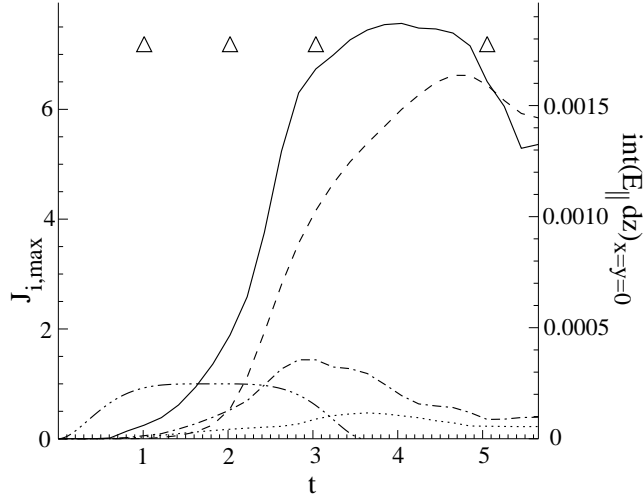}
\caption{Evolution of the maximum value of each component
of ${\bf J}$ ($J_x$ dotted, $J_y$ dot-dashed, $J_z$ solid), as well as the
integral of $E_{\|}$ along the $z$-axis (dashed), and the time-variation of the
boundary driving (triple-dot-dashed). The diamonds ($\bigtriangleup$) mark the
times at which the isosurfaces of $J$ in Fig.~\ref{jisotrans} are plotted.}
\label{jcomptime}
\end{figure}

Now examine the temporal evolution of the maximum value of each component of
the current (see Fig.~\ref{jcomptime}). It is clear that the current component
that is enhanced significantly during the evolution is $J_z$. This is the
component which is parallel to the fan plane, and perpendicular to the plane
of the shear (consistent with Ref.~[\onlinecite{pontingalsgaard2007}], and also with a
collapse of the null's spine and fan towards each other (see, e.g.~Ref.~[\onlinecite{parnell1996}])).

Examining the plasma flow in the plane perpendicular to the shear, ($z$
constant) we find that it develops a
stagnation-point structure, which acts to close up the spine and fan (see
Fig.~\ref{vtransfig}).  It is
worth noting that the Lorentz force acts to accelerate the plasma in this way,
while the plasma pressure force acts against the acceleration of this flow
(opposing the collapse).
In the early stages after the current sheet has formed, the stagnation flow is
still clearly in evidence, with plasma entering the current concentration along its
`long sides' and exiting at the short sides---looking very much like the
standard 2D reconnection picture in this plane
[Fig.~\ref{vtransfig}(b)]. As time goes on, the driving flow in  the
$y$-direction begins to dominate.  

\subsection{Magnetic structure}

\begin{figure}
\centering
(a)
\includegraphics[height=6cm]{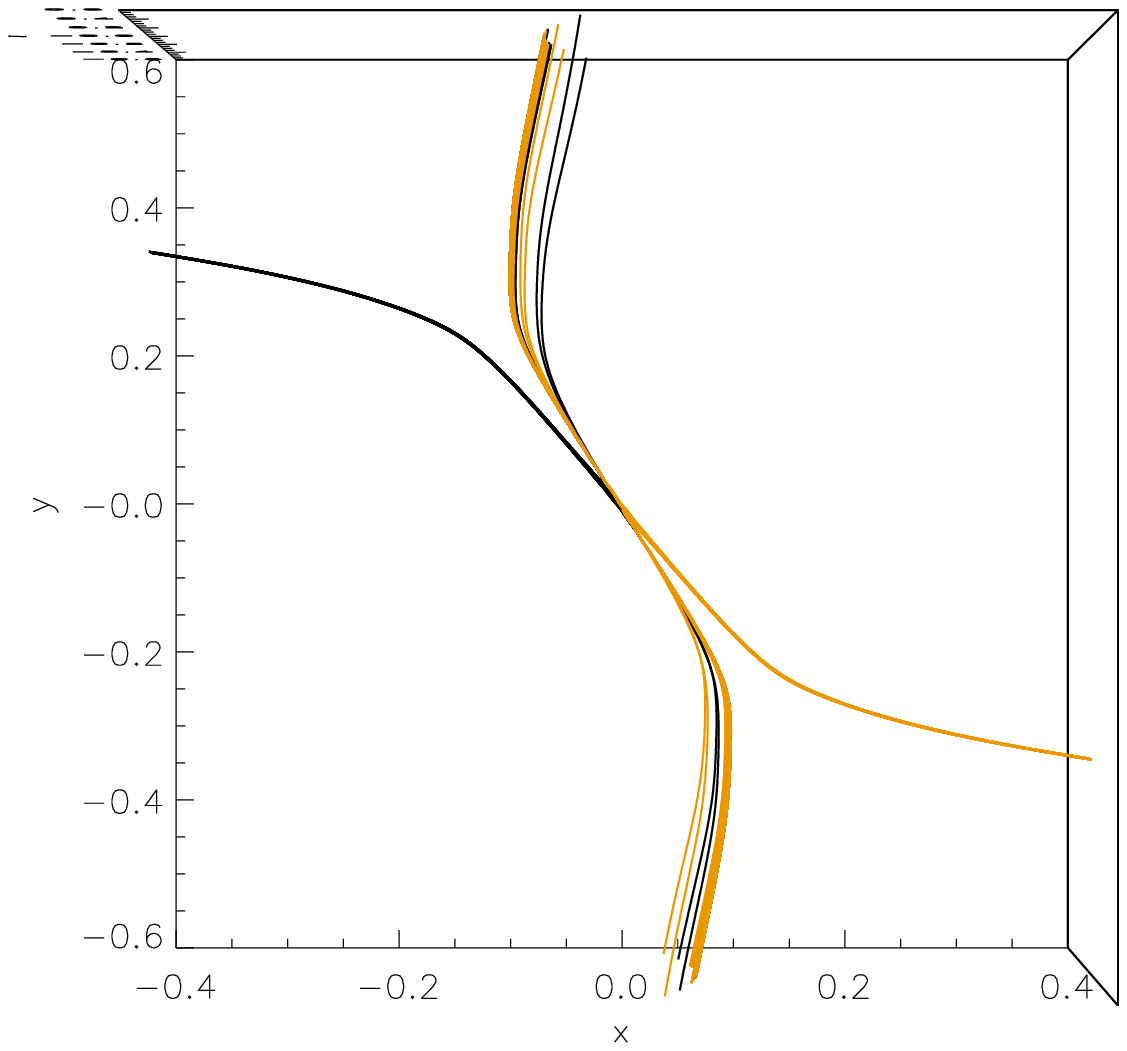}
(b)
\includegraphics[height=6cm]{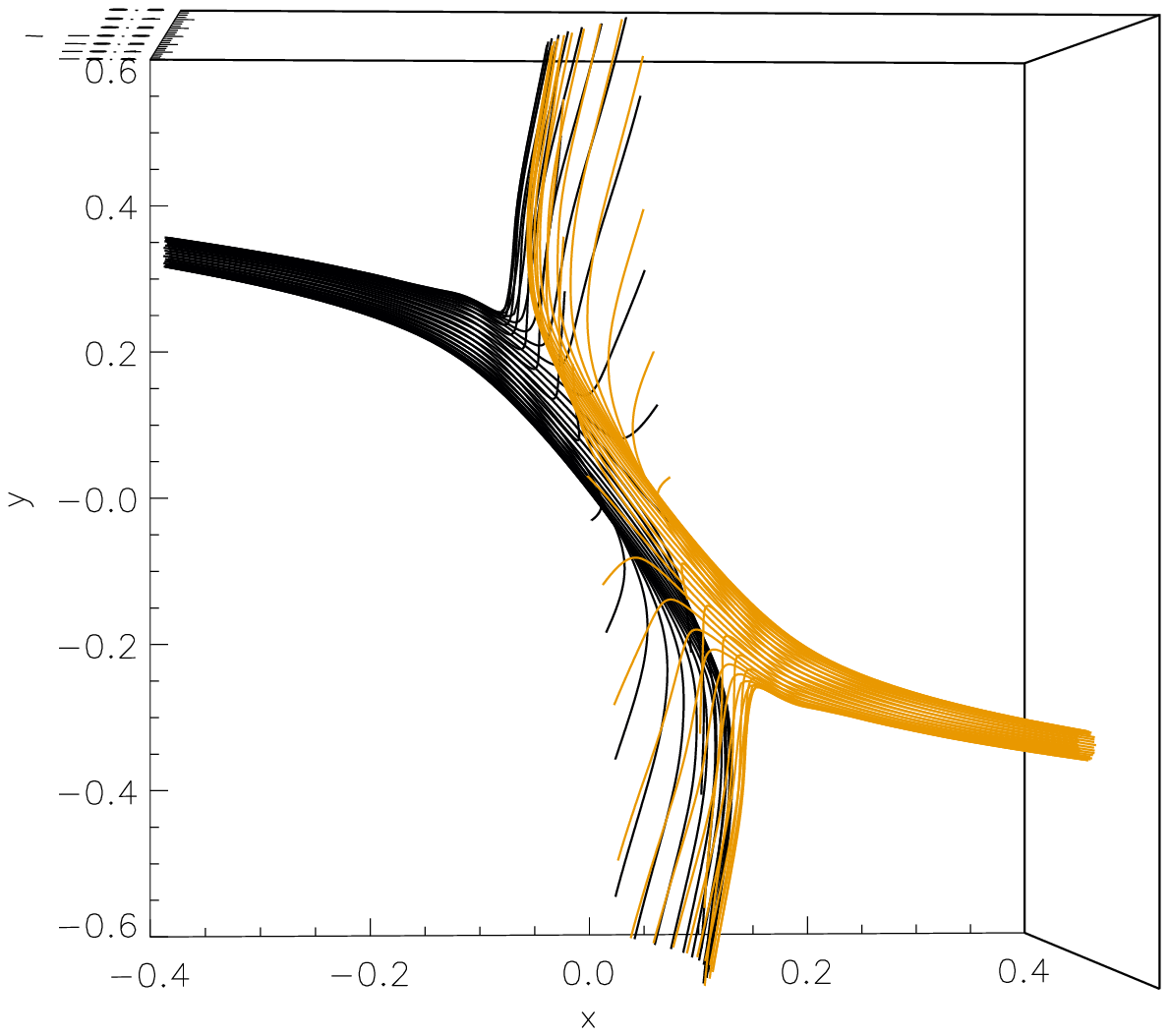}\\
(c)
\includegraphics[height=6cm]{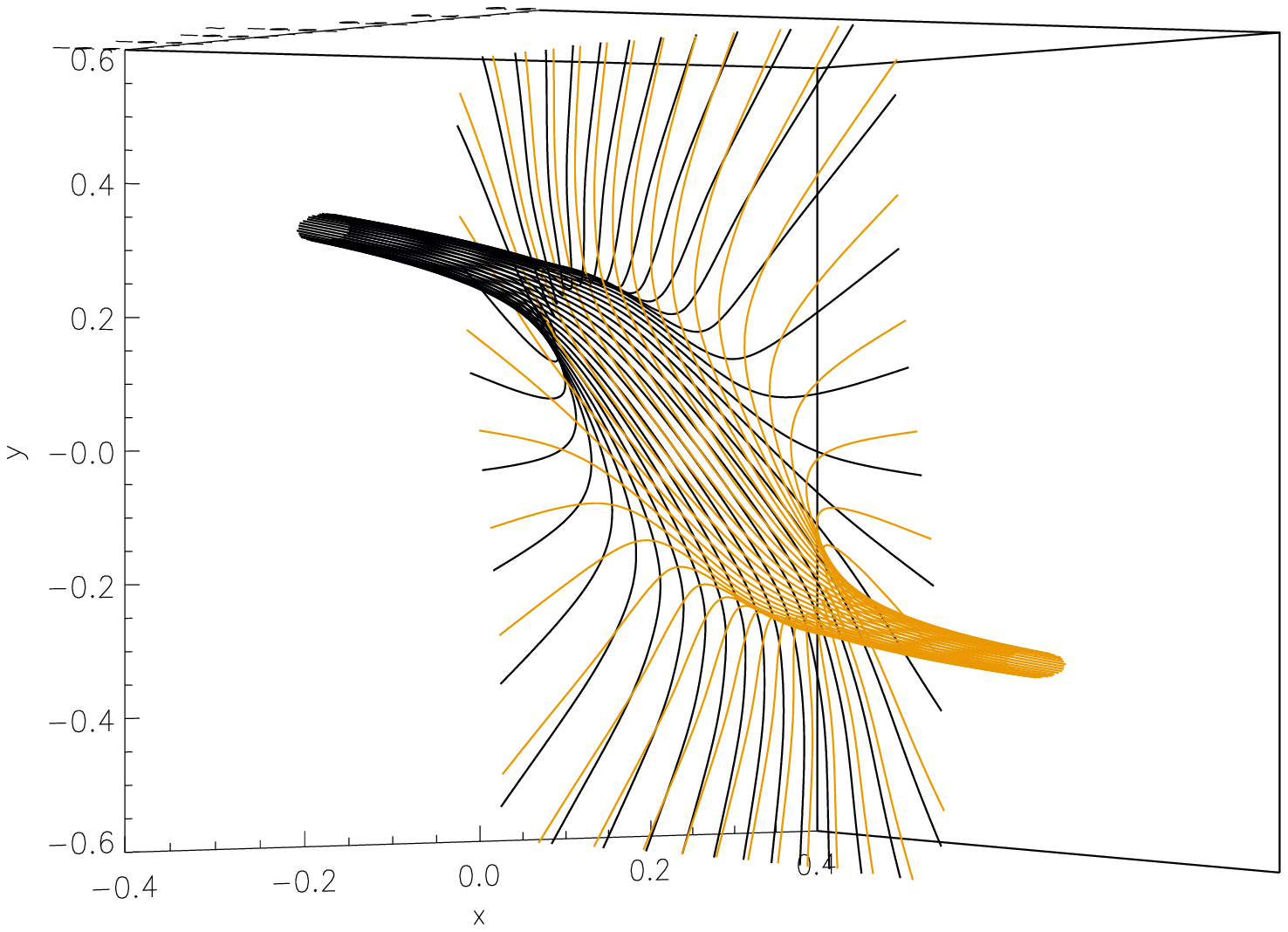}
(d)
\includegraphics[height=6cm]{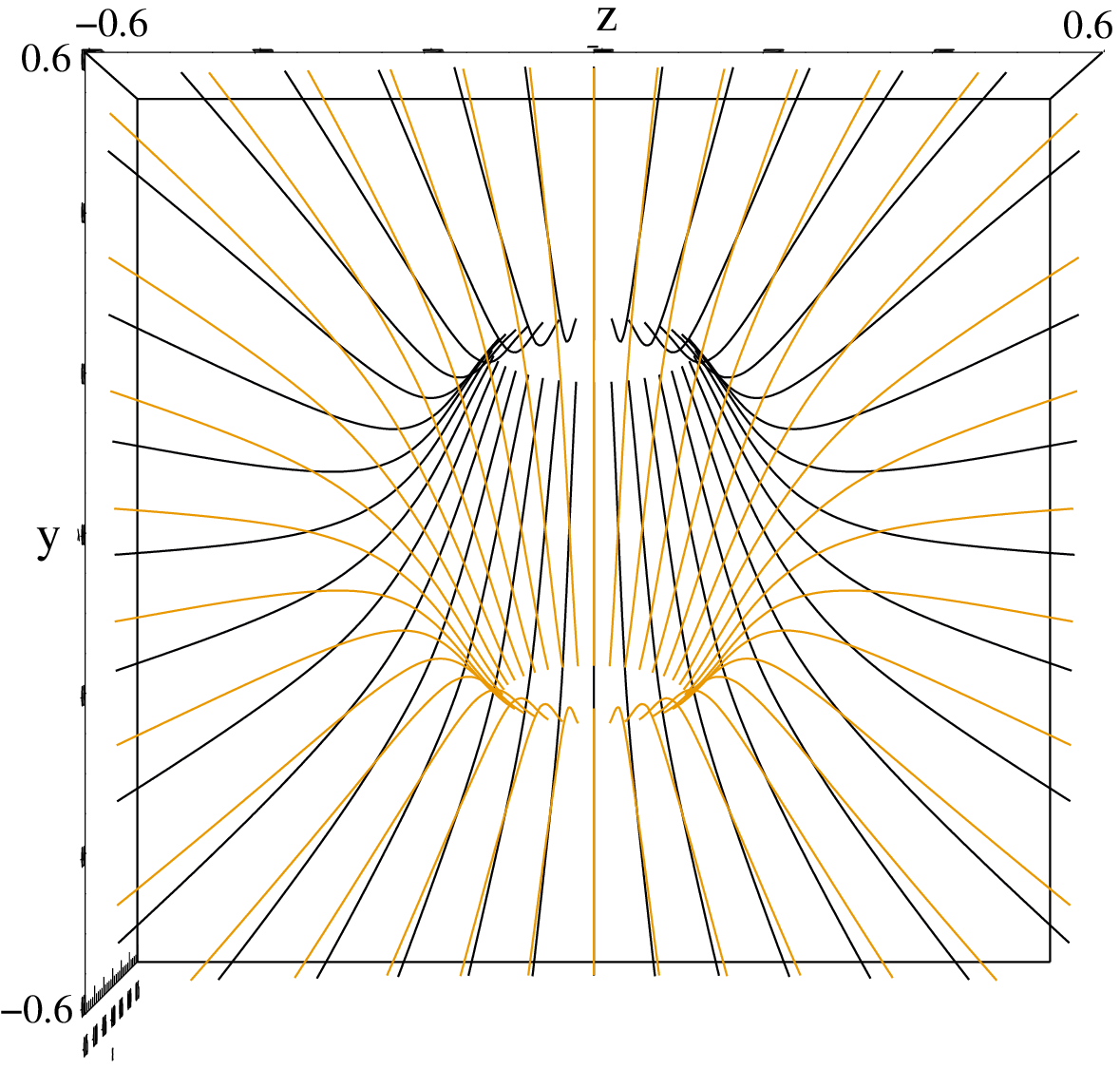}
\caption{(Colour online) (a) Field lines traced from very close to the spine (black for
  negative $x$, grey (orange online) for positive). (b-d) The same, but for field lines
  traced from rings of larger radius around the spine and which graze the
  surface of the current sheet, viewed at different
  angles of rotation about the $y$-axis; (b) $\pi/36$, (c) $\pi/6$, (d)
  $\pi/2$. 
  This simulation run uses continuous driving, with parameters
  $X_l=0.5$, $Y_l=Z_l=6$, $A_d=160$, $v_0=0.04$, $\tau=0.5$ and $\eta=5\times
  10^{-4}$. Time of the images is $t=6$.} 
\label{sheetBstrucfig}
\end{figure}

As the evolution proceeds and the current becomes strongly concentrated at the null,
this is naturally where the magnetic field becomes stressed
and distorted. Firstly, plotting representative field lines which thread the
sheet and pass very close to the null, we see 
that in fact the topological nature of the null is preserved [Fig.~\ref{sheetBstrucfig}(a)]. 
That is, underlying the
Y-type structure of the current sheet there is still only a
single null point present, with the angle between the spine and the fan
drastically reduced (see below).

It is interesting to consider the 3D structure of the current sheet, that is the
nature of the quasi-discontinuity of the magnetic field at the current
sheet. 
This is important since the common conception of a 
`current sheet' is a 2D Green-Syrovatskii  sheet with anti-parallel field on either side of a
cut in the plane (see Refs.~[\onlinecite{green1965,syrovatskii1971}]). Here, however, a current sheet localised in all three dimensions is present.

In order to determine the structure of this 3D current sheet, we would like to know the nature of the jump ($\Delta {\bf B}$) in the magnetic field vector ${\bf B}$ across it, and how this varies throughout the sheet. Of course, because $\eta\neq 0$, there is no true discontinuity in ${\bf B}$, though the jump $\Delta {\bf B}$ would occur over a shorter and shorter distance if $\eta$ were decreased (in an external domain of fixed size). This mismatching in ${\bf B}$ may be visualised by plotting field lines which define approximately the boundaries of the
current sheet, see Fig.~\ref{sheetBstrucfig}(b-d). Along  $z=0$ between the two ends of the current sheet, the black and grey (orange online) field
lines (on opposite sides of the current concentration) are exactly anti-parallel [see Fig.~\ref{sheetBstrucfig}(d)]. Thus in this plane (only) we have something similar to the 2D picture (due to symmetry). However, for $z\neq 0$, the magnetic field vectors on either side of the current sheet are not exactly anti-parallel, and the black and grey (orange online) field lines cross at a finite angle (for small $|y|$). This angle decreases as $|z|$ increases (and as $|y|$ increases for $z\neq 0$), and thus the current modulus---proportional to $\Delta {\bf B}$ across the sheet---falls off in $y$ and $z$ [see Fig.~\ref{sheetBstrucfig}(d)], and is localised in all three dimensions [compare Fig.~\ref{jisotrans}(c) and Fig.~\ref{sheetBstrucfig}(c)].


\subsection{Eigenvectors and eigenvalues}
\begin{figure}
\centering
\includegraphics{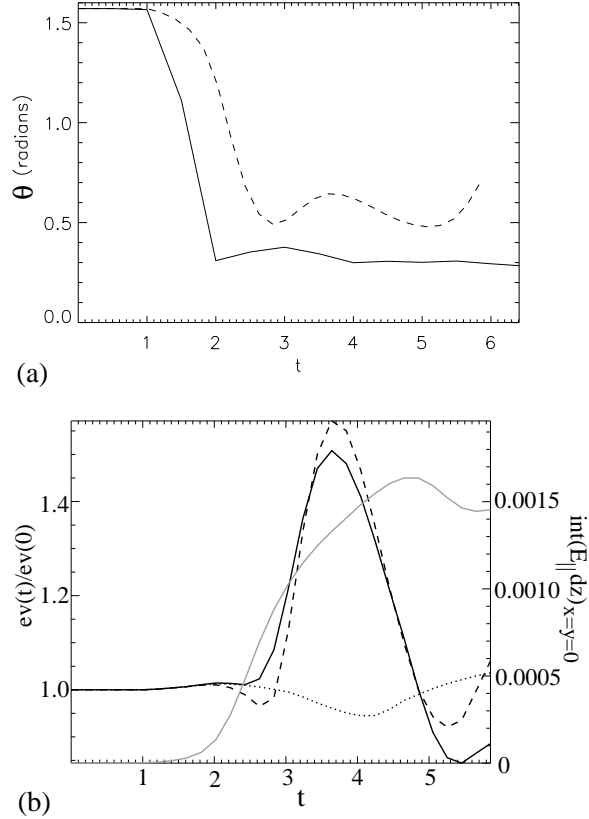}
\caption{(a) Evolution of the angle ($\theta$) between the spine and fan (eigenvectors), for
  continual driving with parameters as in Fig.~\ref{sheetBstrucfig} (solid line) and
  the transient driving run (dashed). (b) Evolution of the null point eigenvalues
  ($x$-eigenvalue solid line, $y$ dashed, $z$ dotted),
  and the integrated parallel electric field along the $z$-axis (grey) for the
  transient driving run. The
  eigenvalues are normalised to their values at $t=0$, for clarity.}
\label{eigangle}
\end{figure}

Examining the evolution of the eigenvalues and eigenvectors of the null point in
time provides an insight into the changing structure of the null. In order to
simplify the discussion, we refer to the eigenvectors which lie along the $x$, $y$,
and $z$ axes at $t=0$ as the $x$-, $y$- and $z$-eigenvectors, respectively, and
similarly for the eigenvalues. Consider
first the eigenvectors. The orientation of the $z$-eigenvector is essentially
unchanged, due to the symmetry of the 
driving. However, the $x$- and $y$-eigenvectors do change their orientations
in time. In
Fig.~\ref{eigangle}(a) the angle ($\theta$) between these two eigenvectors, or equivalently
between the spine and fan, is plotted in time. In the case of continuous
driving, the initial angle of $\pi/2$ quickly closes
up to a basically 
constant value once the sheet forms. With the transient driving, the pattern
is the same, except that the angle starts to grow again once the driving
switches off (note that the minimum angle is smaller for the continually
driven run plotted, since the driving $v_0$ is 4 times larger).

The change in time of the eigenvalues [see Fig.~\ref{eigangle}(b)], and their ratios, is
linked to that of the eigenvectors. The
eigenvalues stay constant in time during the early evolution, but then begin
to change, although they each follow the same pattern (i.e.~they maintain
their ratios, between approximately $t=1$ and $t=2$). By contrast, once the null
point begins to 
collapse, there is a significant time dependence to the eigenvalues ($t>2$), and
evidently also to their ratios. This is clear evidence of non-ideal behaviour at
the null, as 
described in Sec.~\ref{kinematic}, and demonstrates that the evolution
modelled by the kinematic example given there is in fact a natural one. That
is, the evolution prohibited under the restriction of ideal conditions is
precisely that which occurs when the null point experiences a typical
perturbation.


\subsection{Parallel electric field and reconnection}
Further indications that non-ideal processes are important at the null are
given by the presence of a component of ${\bf E}$ parallel to ${\bf B}$
($E_{\|}$) which develops there. The presence of a current sheet at the null,
together with a parallel electric field, is a strong indication that
reconnection is taking place in the vicinity of the null (we actually expect that field lines 
change their connectivity everywhere within the volume defined by the
current sheet (the `diffusion region'), not only at the null itself \cite{priesthornig2003,
  pontinhornig2005}, which is special only in that the footpoint mapping is discontinuous there). 

\begin{figure}
\centering
\includegraphics[scale=1.2]{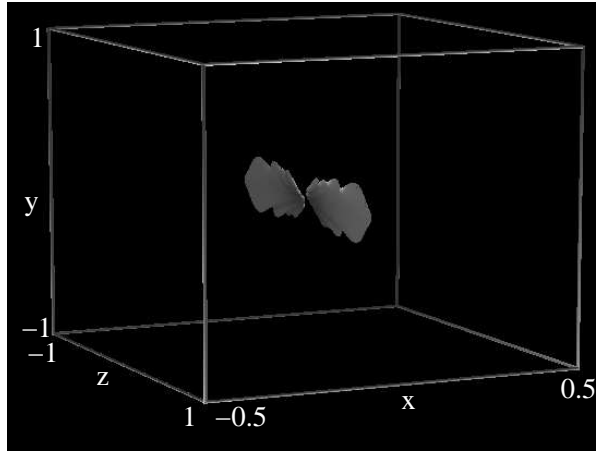}
\caption{Isosurface of $E_{\|}$ at $t=4.75$ (time of its maximum), at
  $65\%$ of maximum, transient driving run.}
\label{isoepar}
\end{figure}

It can be shown \citep{pontinhornig2005} that $\Psi=\displaystyle \int_{x=y=0}
E_{\|} \: dz$ provides a measure of the reconnection  rate at the null. 
  This quantity gives an exact measure of the
  rate of flux transfer across the fan surface when the non-ideal region is
  localised around the null point.
The spatial distribution of $E_{\|}$ within the domain when it
reaches its temporal maximum is closely focused around
the $z$-axis (see Fig.~\ref{isoepar}). This is because ${\bf J} \| {\bf B}$
there by symmetry (note also that $E_{\|}$ is discontinuous at the plane $z=0$
by symmetry, since $B_z$ changes sign through this plane but $J_z$ (and thus
$E_z$) is  uni-directional---however, $\displaystyle\int E_{\|} \:ds$ is
non-zero since $ds$ also changes 
sign through $z=0$). The field line which is coincident with the $z$-axis (by 
symmetry) thus provides the maximum value for $\displaystyle \int E_{\|} \:
ds$ of any field line threading the current sheet---another reason to
associate this quantity with the reconnection rate. 

\begin{figure}
\centering
(a)\includegraphics[scale=0.37]{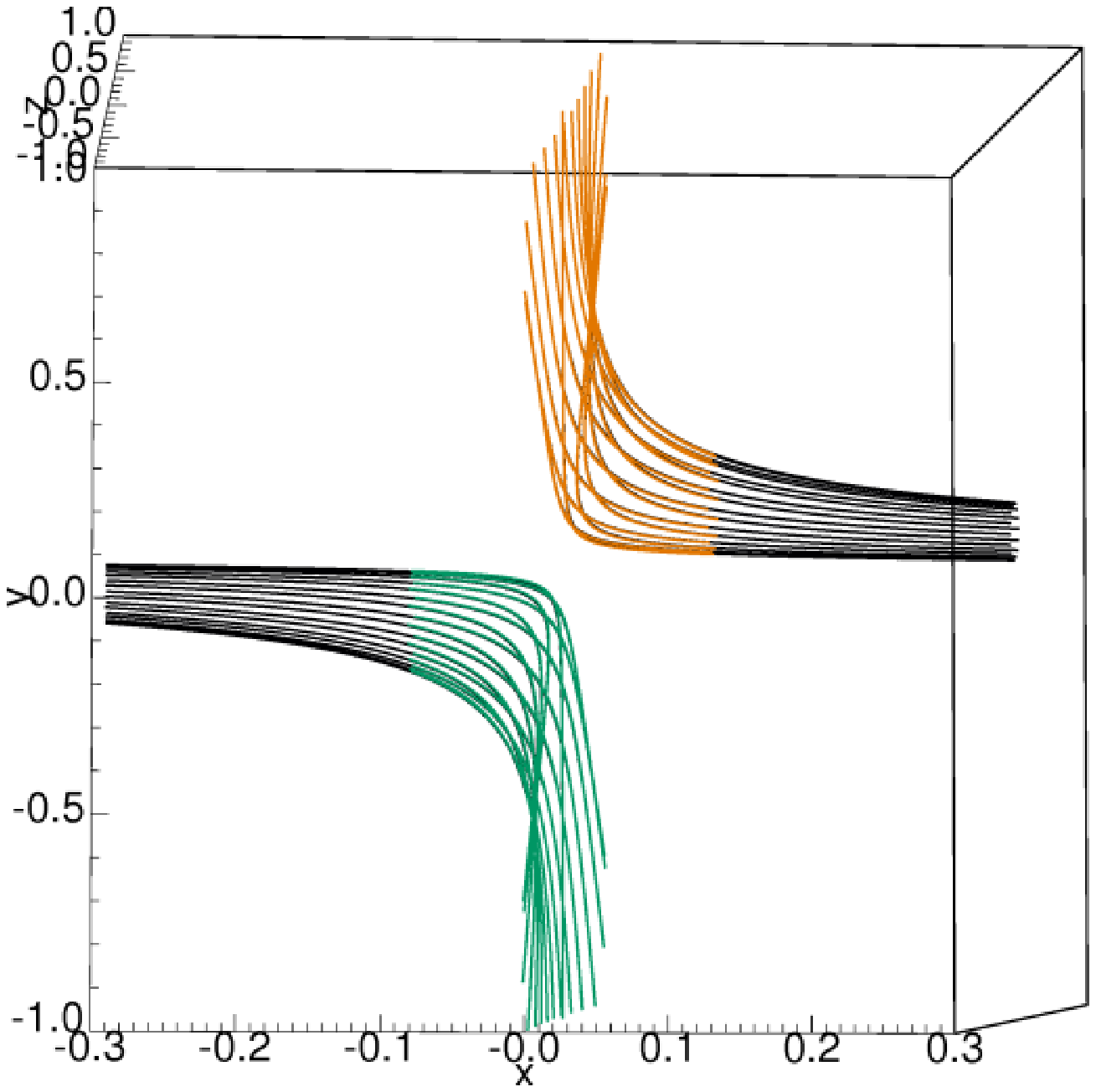}
(b)\includegraphics[scale=0.37]{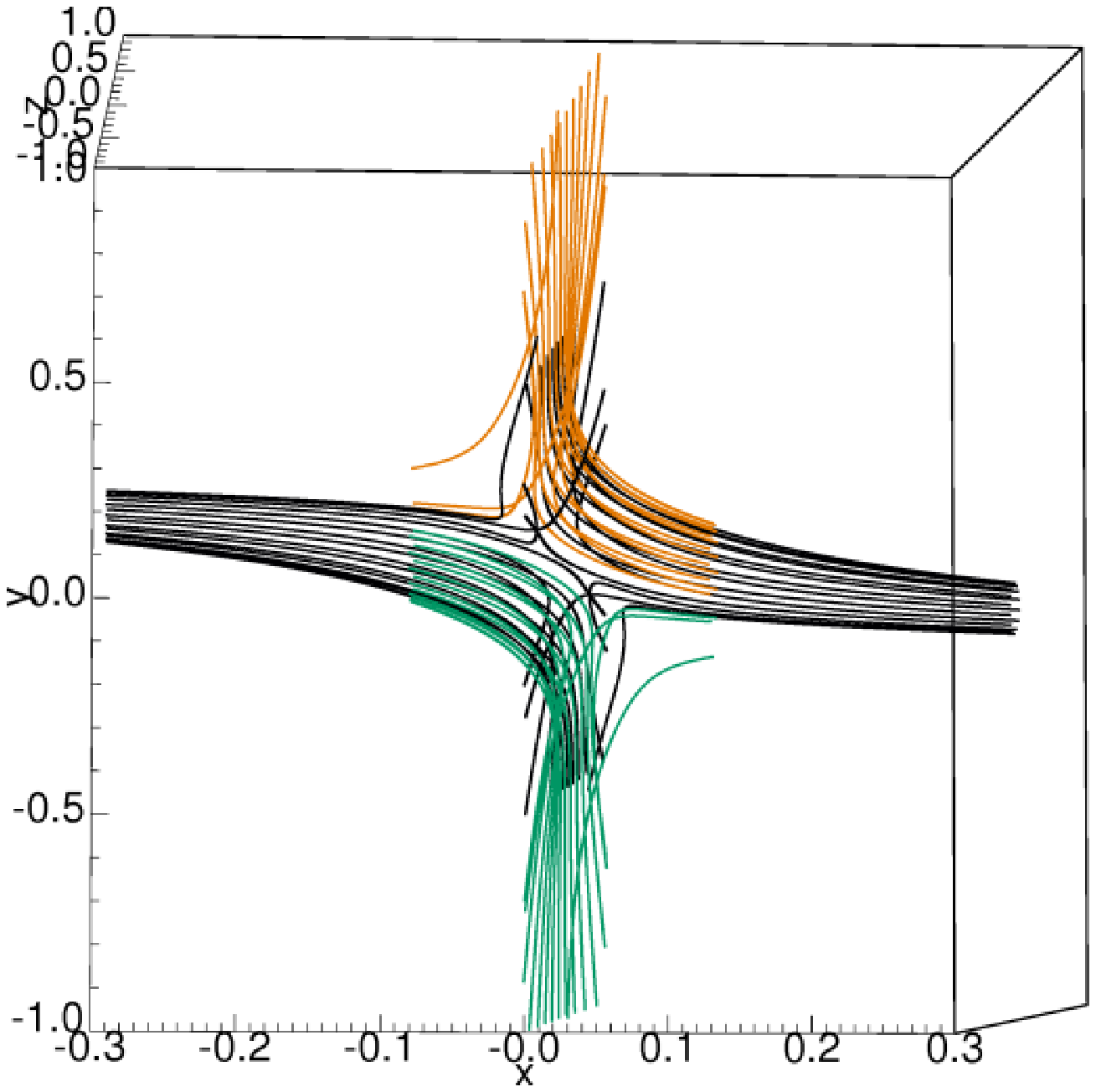}
(c)\includegraphics[scale=0.37]{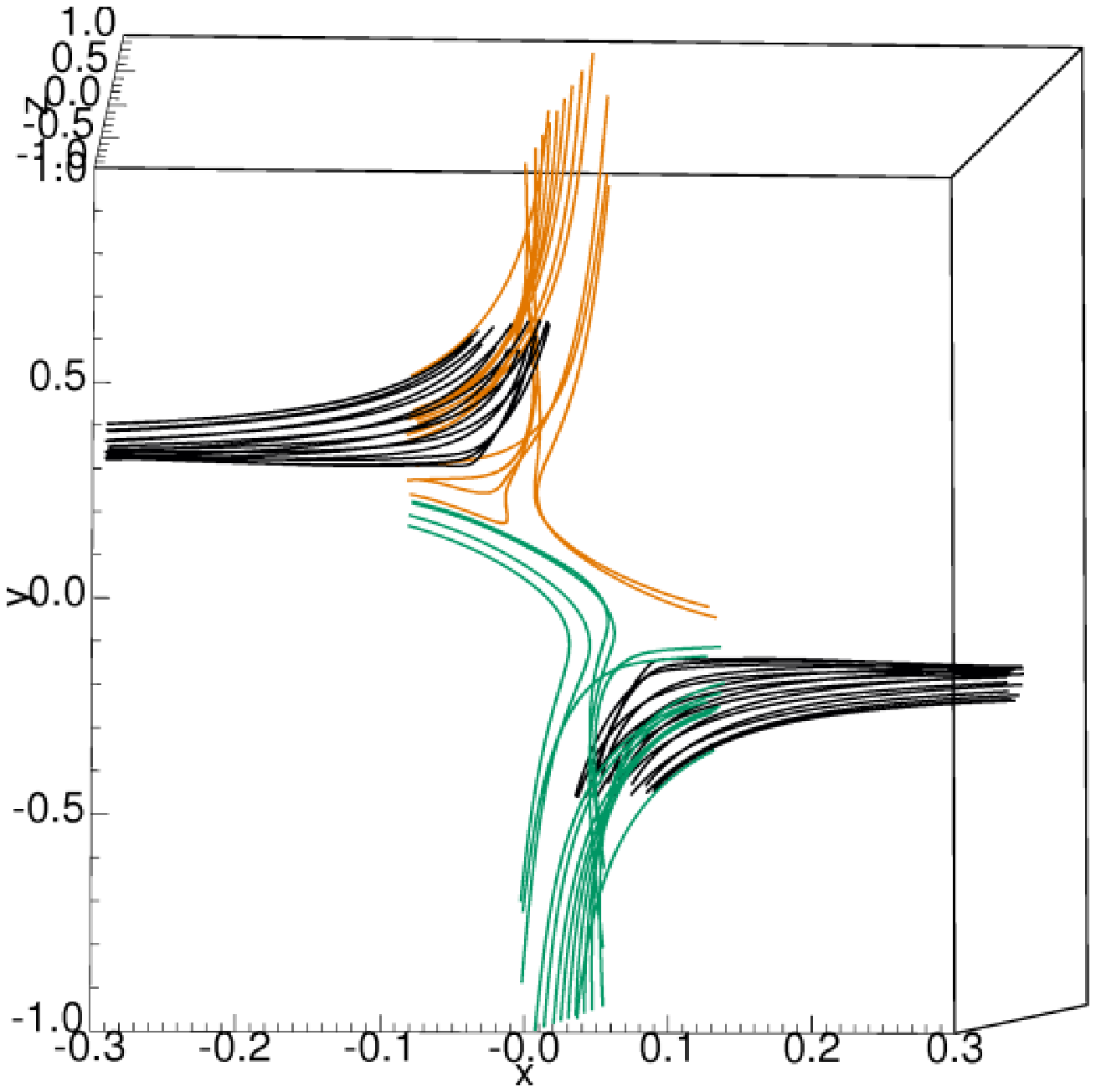}
\caption{(Colour online) Field lines integrated from ideal trace particles, located initially
  far out along the spine (black) and the fan (grey, green and orange
  online). Images for (a) $t=0$, (b) $t=3$, (c) $t=6$, and for continual
  driving with parameters as in Fig.~\ref{sheetBstrucfig}, but $v_0=0.03$.}
\label{blinerec}
\end{figure}

The nature of the field line reconnection can be studied by integrating field
lines threading `trace particles', which move in the ideal flow far from
the current sheet, and which initially define coincident sets of field
lines (see Fig.~\ref{blinerec}). Once the current sheet forms, field lines
traced from particles far out along the spine and far out along the fan are no
longer coincident (i.e.~they have been `reconnected', near the null). Field
lines traced from around the spine flip around the spine, while
there is also clearly advection of magnetic flux across the fan
surface. However, it seems that this occurs in the main
part during the collapse of the null, while later in the simulations
reconnection around/through the spine is dominant, since the boundary driving
is across the spine. 



\section{Quantitative current sheet properties}\label{quantsec}
In order to understand the nature of the current sheet that forms at the null
point, its quantitative properties should be analysed. We focus here on two
main aspects, firstly the scaling of the current sheet with the driving
velocity, and secondly its behaviour at large times under continual driving. 
Crucial in this is whether the behaviour tends towards that of a
Sweet-Parker-type current sheet, that is whether the sheet's length increases eventually to
system-size, and whether the peak reconnection rate scales as a negative
power of $\eta$, thus providing only slow reconnection for small $\eta$. 
While
the computational cost of a scaling study of $J_{max}$ with $\eta$ for a fully
3D simulation such as ours is prohibitive, the investigations which we do
perform provide an indication of the nature of the sheet.

\subsection{Scaling with $v_0$}\label{v0scale}
Firstly we consider the scaling of the current sheet with 
the magnitude of the boundary driving velocity.
In particular, we look at the maximum current density which is attained
(which invariably occurs at the null), the maximum
reconnection rate (calculated as the integrated parallel electric field along
the fan field line coincident with the $z$-axis, as described previously), and
also the current sheet dimensions $L_x$, $L_y$ and $L_z$ (taken to be the full
width at half maximum in each coordinate direction). This is done for the case
of transient driving, with $\tau$ fixed at a value of $1.8$.
\begin{figure}
\centering
\includegraphics[scale=1.2]{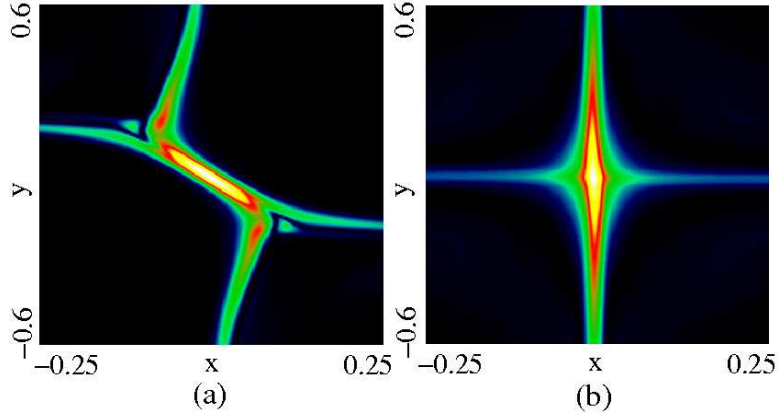}
\caption{(Colour online) Current modulus in the $z=0$ plane at time of current maximum, for (a)
  $v_0=0.04$, (b) $v_0=0.001$, scaled in each case to the individual
  maximum (20.4 in (a) and 0.44 in (b)).} 
\label{jsheetstrongdr}
\end{figure}
One point which should be taken into account when considering the measurements
of the dimensions
of the current sheet (particularly in the $x$- and $y$-directions), is that
the measurements do not necessarily mean exactly the same thing in the
different simulation runs, in the sense that the current sheet morphology is not
always qualitatively the same. Observe the difference between
the current structures in Fig.~\ref{jsheetstrongdr}, which shows cases with
different driving strengths. 
When the driving in stronger, the current sheet is more
strongly focused, and we measure a straight current
sheet, which spans the spine and fan [Fig.~\ref{jsheetstrongdr}(a)]. 
However, when the driving is weaker, the region of high (above one half
maximum) current density spreads along the fan plane, in an `S' shape in and $xy$-plane
[Fig.~\ref{vtransfig}(c)]. Finally, when $v_0$ is decreased further
still, we again have an approximately planar current sheet, but this time lying
in the fan plane [Fig.~\ref{jsheetstrongdr}(b)]. Note that this changing
morphology is also affected by the parameters $\eta$, $\beta$ and $\gamma$,
which we will discuss in a future paper, but which are held fixed here.

\begin{figure}
\centering
\includegraphics[scale=1.2]{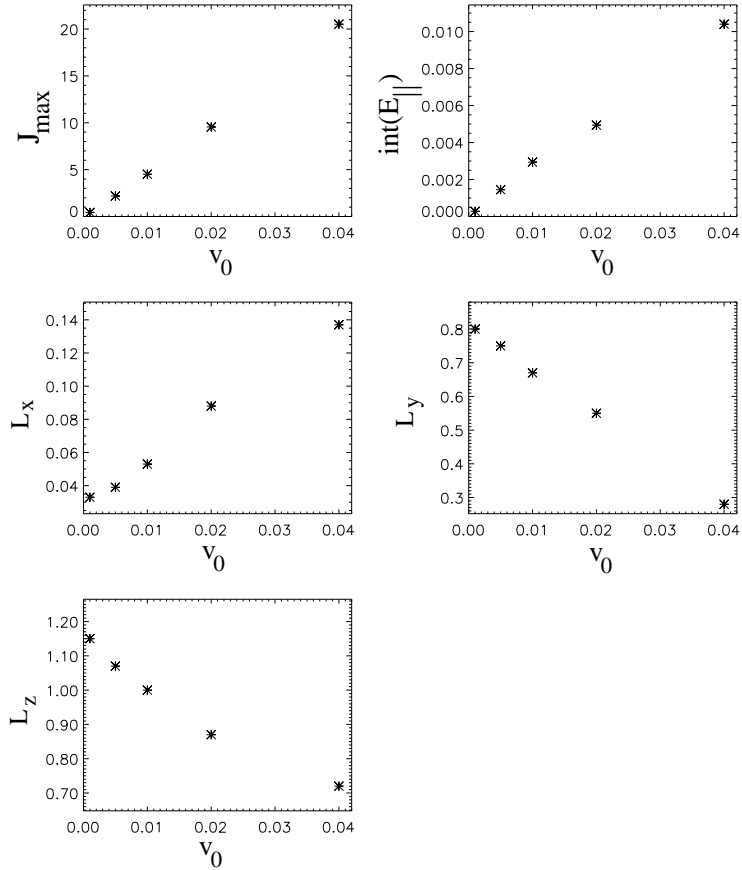}
\caption{Scaling  with the modulus of the driving velocity ($v_0$) of the
  peak current ($J_{max}$), the peak reconnection rate ($\int E_{\|}$), and
  the full width at half maximum of the current sheet in each coordinate
  direction ($L_x, L_y, L_z$).}
\label{jscale-drvel1}
\end{figure}
We repeat the simulations with transient driving, varying $v_0$, but fixing 
$X_l=0.5$, $Y_l=Z_l=3$, $A_d=80$, $\tau=1.8$ and $\eta=5\times 10^{-4}$.
The results are shown in Fig.~\ref{jscale-drvel1}.
First, we see that the peak current and peak reconnection
rate both scale linearly with $v_0$. The extent of the current sheet in $x$
($L_x$) increases linearly with 
$v_0$, which is a signature of the increased collapse of the null. 
By contrast, $L_y$ decreases with increased driving
velocity. This is rather curious and seems counter-intuitive. It appears that
the current sheet is more intense and more strongly focused at the null for
stronger driving. In addition though, it is a result of the fact that an
increasing amount of
the current is able to be taken up in a straight spine-fan spanning current
sheet, rather than spreading in the fan plane.
Likewise the scaling of $L_z$ is also curious---the current sheet is more
intense and strongly focused at the null for stronger driving.

The above scaling analysis has also been performed for the case where the
spine is displaced by the same amount each time, but at different rates. That
is, as $v_0$ is increased, $\tau$ is decreased to compensate. In fact the
scaling results are very similar to 
those above, but just with a slightly weaker dependence on $v_0$.

\subsection{Long-time growth under continual driving}
Now consider the case where, rather than imposing the boundary driving for
only a limited time, the driving velocity is ramped up and then held
constant. 
We take parameters as in Sec.~\ref{curevsec}, but with $Y_l=Z_l=6$ and run resolution
$128\times192\times192$, giving minimum $\delta x \sim0.005$ and $\delta y, \delta z \sim 0.025$ again at the null.
We focus on whether the current sheet continues to grow when it
is continually driven, or whether it reaches a fixed length due to some
self-limiting mechanism. 
As before, one of the major issues we run into
between different simulation runs which use different parameters is in the
geometry of the current sheet. For simplicity here we
consider a case with sufficiently strong driving that the current sheet is
approximately straight, spanning the spine and fan, as in
Fig.~\ref{jsheetstrongdr}(a). 

\begin{figure}
\centering
(a)\includegraphics[height=5cm]{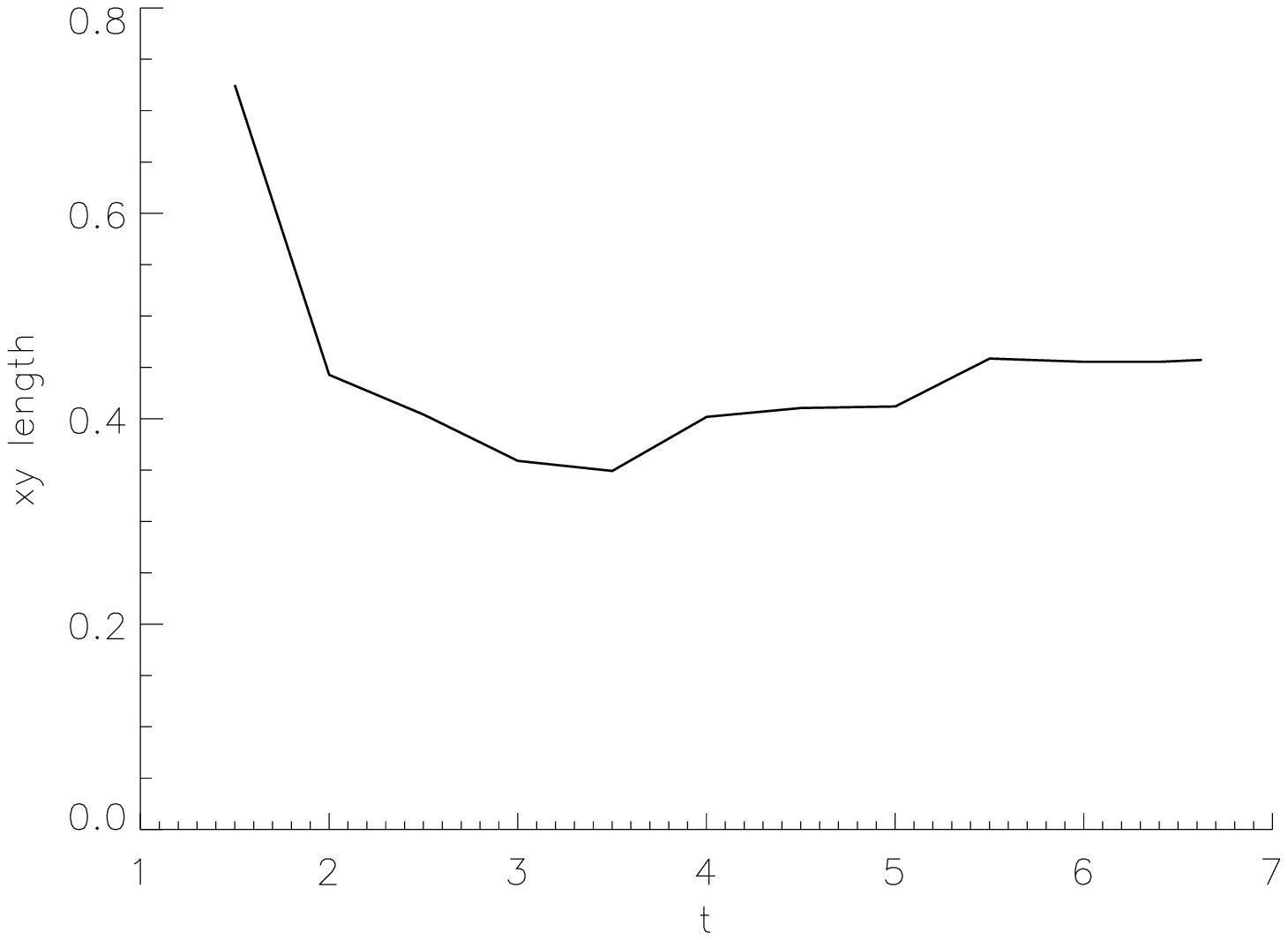}
(b)\includegraphics[height=5cm]{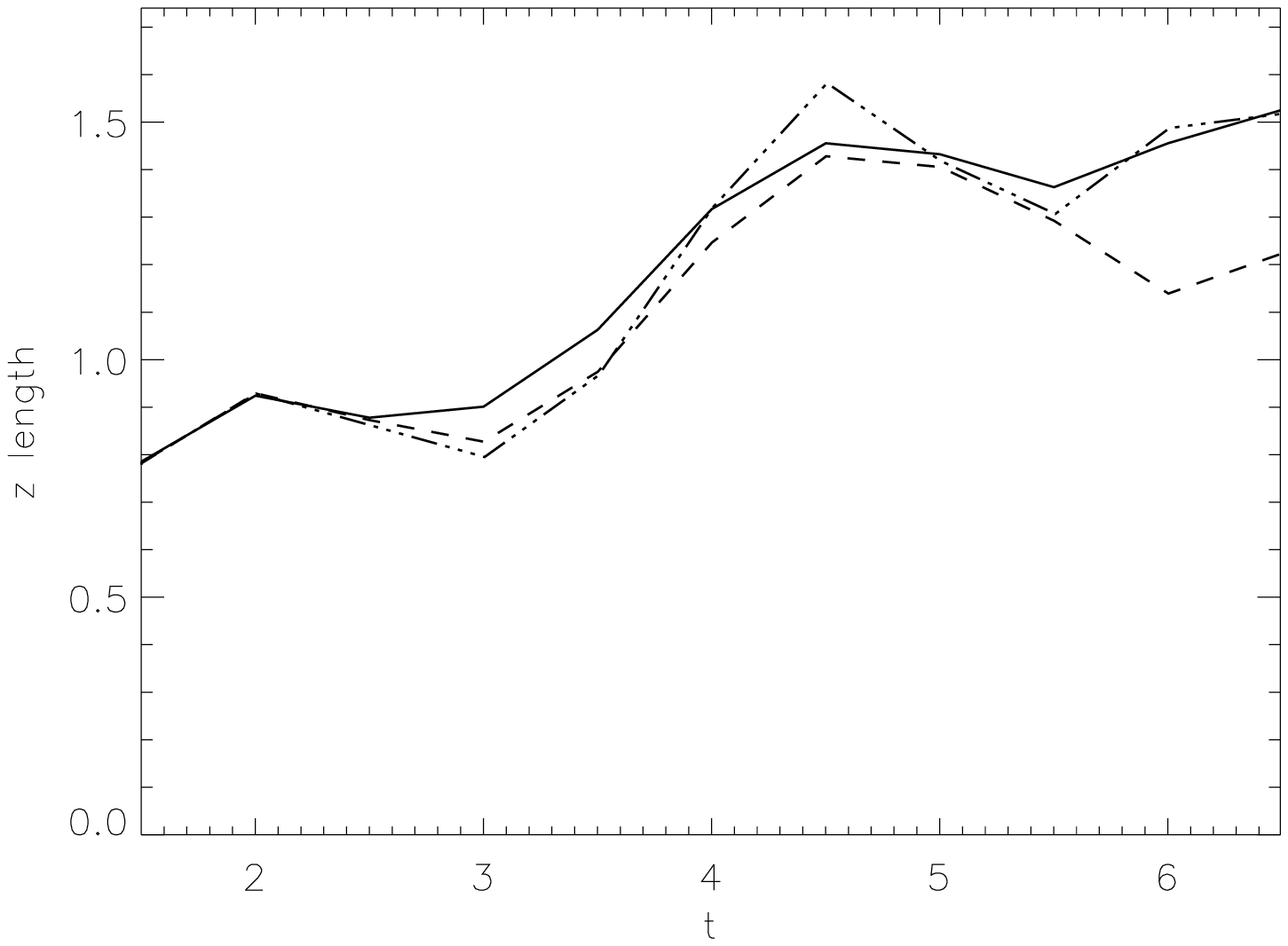}
(c)\includegraphics[height=5cm]{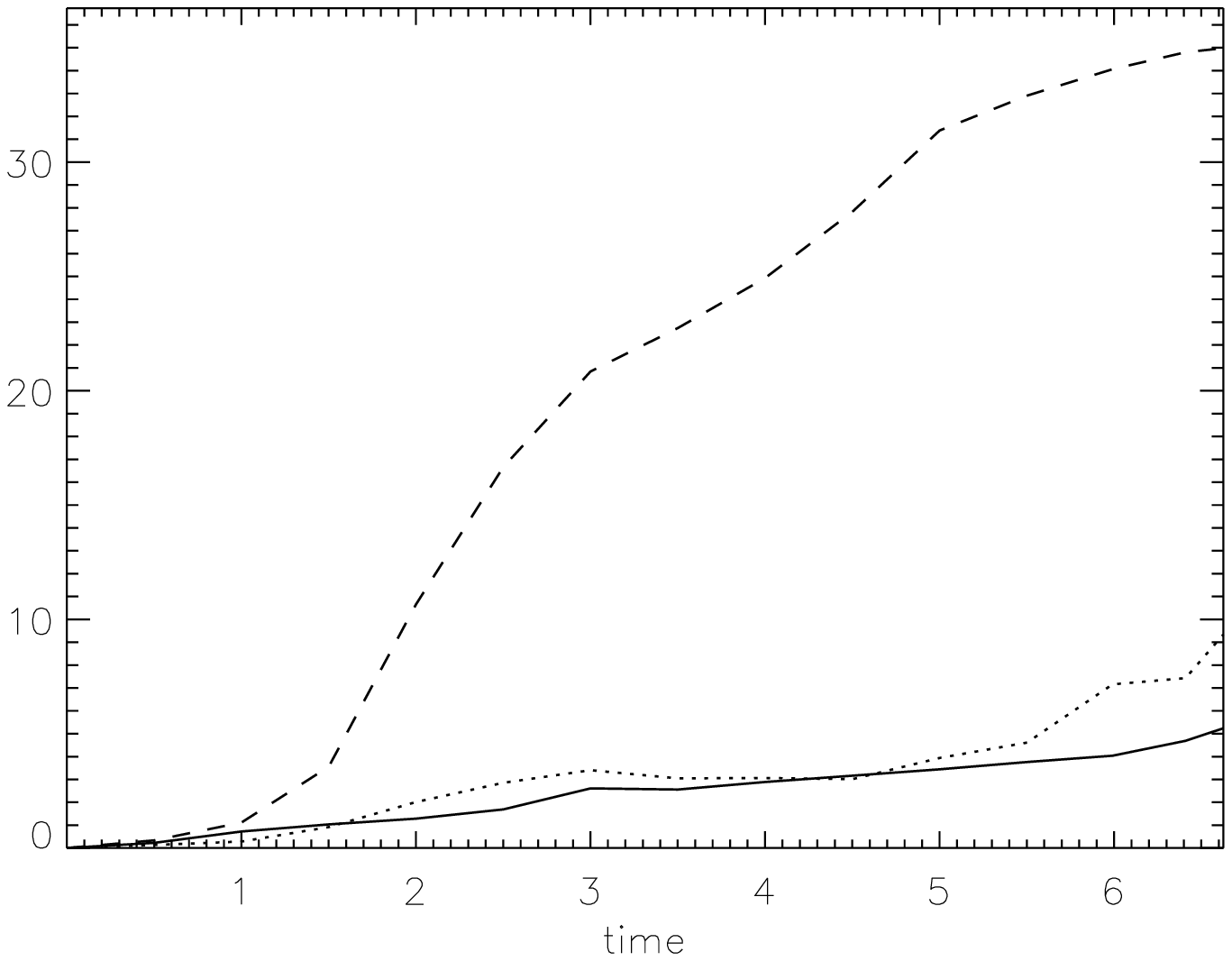}
\caption{Growth of the sheet in time, in (a) the
  $xy$-direction ($\sqrt{L_x^2+L_y^2}$) and (b) the $z$-direction for three runs 
  with different domain sizes ($Y_l=Z_l=6$ solid line,
  $Y_l=Z_l=3$ dashed, $Y_l=Z_l=1.5$ dot-dashed). (c) Time evolution of the
  maximum value of each current component ($J_x$ solid line, $J_y$
  dotted, $J_z$ dashed.)}
\label{sheetlent}
\end{figure}

There are many problems which make it hard to determine the
time evolution of the current sheet length.
During the initial stage of the sheet formation, it actually
shrinks, as it intensifies, in the $xy$-plane (measuring the length as
$\sqrt{L_x^2+L_y^2}$, see Fig.~\ref{sheetlent}(a)). After this has occurred, the sheet 
then grows very slowly---differences over an Alfv{\' e}n time are on
the order of the gridscale. There is no sign though of any evidence which
points to a halting of this growth. Similarly, examining the evolution of $L_z$,
it seems to continue increasing for as long as we can run the
simulations. Neither of the above is controlled by the dimensions of the
numerical domain (see Fig.~\ref{sheetlent}(b)).

As to the value of the peak current, and the reconnection rate at the
null, though the growth of these slows significantly as time progresses (see
Fig.~\ref{sheetlent}(c)), there is again no sign of them reaching a saturated
value. The slowing of the current sheet growth (dimensions and modulus) is
undoubtedly down to the diffusion and reconnection which occurs in the current
sheet. 

In summary, it is difficult to be definitive that the current sheet will grow to system-size under continuous driving, due to the computational difficulties of running for a long
time. However, there is no indication that the dimensions of the current sheet are limited by any self-regulating process. Rather, the dimensions are determined by the boundary conditions (i.e.~the degree of shear imposed from the driving boundaries at a given time). Thus, if it were computationally possible to continue the shearing indefinitely, it appears that the sheet would continue to grow in size and intensity. If the system size is large and the resistivity is low, it is possible that the extended current sheet may break up into secondary islands, which lie beyond the scope of the present simulations.

\section{Possibility of turbulent reconnection}\label{turbsec}
\begin{figure}
\centering
\includegraphics[scale=1.2]{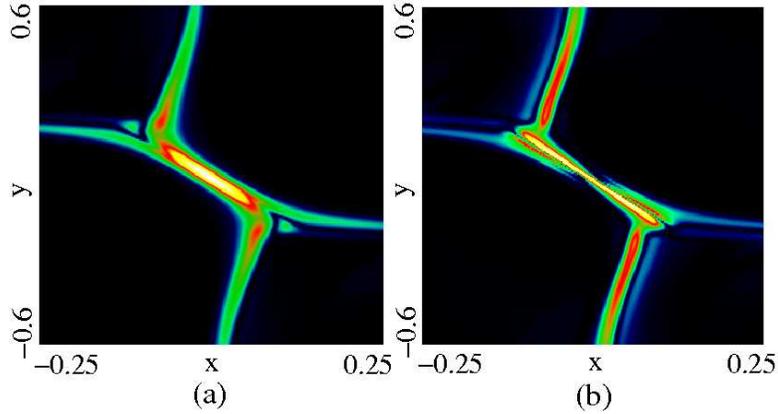}
\caption{(Colour online) (a) Current and (b) vorticity profiles in the $z=0$ plane through the null,
  at the time of peak current, for parameters as in Sec.~\ref{v0scale}, with
  $v_0=0.04$. }
\label{curvort}
\end{figure}

A crucial aspect of any reconnection model which hopes to explain fast energy
release is the scaling of the reconnection rate with the dissipation
parameter. Mechanisms for turbulent reconnection have been put forward which
predict reconnection rates which are completely independent of the resistivity
\cite{lazarian1999}. Recent work by Eyink and Aluie \cite{eyink2006}, who obtain
conditions under which Alfv{\' e}n's ``frozen flux'' theorem may be violated in
ideal plasmas, has placed rigourous constraints on models of turbulent
reconnection. The breakdown of Alfv{\' e}n's theorem occurs over some length scale
defined by the turbulence, which may be much larger than typical dissipative
length scales. Eyink and Aluie demonstrate that in such a turbulent plasma, a necessary condition for such
a breakdown to occur is that current and vortex sheets intersect one another
\footnote{In addition to the necessary condition stated above, Eyink and Aluie
  suggest 
two other possible necessary conditions: one, when the advected loops are
non-rectifiable, and two, when the velocity or magnetic fields are unbounded.
These latter two conditions appear to us to be less physically interesting in
the present context.}.
This is a rather strong condition on the nonlinear dynamics underlying a
turbulent MHD plasma. (Note that in order to correspond to any type of
reconnection geometry, more than two current/vortex sheets should intersect,
or equivalently they should each have multiple `branches', as in
Fig.~\ref{curvort}, say along the 
separatrices. This ensures a non-zero electric field at the
intersection point, unlike in the simplified example of Eyink and Aluie.)

One possible viewpoint of how such a situation might occur is that these
current sheets and vortex sheets are generated by the turbulent mechanism
itself. Alternatively, one might imagine another situation in which the result
might be applicable is in a configuration where macro-scale current sheets and
vortex sheets are present in the laminar solution, which may then be modulated
by the presence of turbulence. The results discussed in the
previous sections point towards 3D null points as sites where this might
occur.

Firstly, examining the vorticity (${\bf \omega}$) profile in the simulations,
we find that in  fact a highly localised region of strong vorticity is indeed present.
Furthermore, this region intersects with the region of high current
density (see Fig.~\ref{curvort}) and is focused at the null (and possibly
spread along the fan surface as described above, depending on the choice of
parameters).
It also appears from the figure that in fact the vorticity profile forms in
narrower layers than the current (in the $xy$-plane), one focused along the
spine, with a much stronger layer in the fan (while the current sheet spans
the spine and fan), so that rather than being completely coincident, the
${\bf J}$ and ${\bf \omega}$ sheets really do `intersect'. The rates at which
${\bf J}$ and ${\bf \omega}$ fall off in $z$ away from the null are also very
similar. 

Moreover, we have seen in Sec.~\ref{kinex} that in order 
for a null point to evolve in certain ways, non-ideal processes are required.
When the ideal system is supposed to evolve in such a way, a
non-smooth velocity profile results, which typically shows up along either
the spine or fan of the null, or both, depending on the boundary
conditions. This non-smooth velocity corresponds to a singular vortex
sheet. 

Finally, it is worth noting that in non-laminar magnetic fields, 3D nulls are
expected to cluster together, creating `bunches' of nulls \cite{albright1999}, thus providing the
possibility that multiple coincident current and vortex sheets might be
closely concentrated.


\section{Summary}\label{conc}
We have investigated the nature of the MHD evolution, and current sheet
formation, at 3D magnetic nulls. In complex 3D fields, isolated nulls are
often considered less important sites of energetic phenomena than separator
lines, due to the viewpoint that ideal MHD singularities in kinematic analyses
are a result of the choice of boundary conditions. However, as demonstrated
here (and proven in Ref.~[\onlinecite{hornig1996}]), certain evolutions of a 3D null
are prohibited under ideal MHD, specifically those which correspond to a time
dependence in the eigenvalue ratios of the null.  We presented a particular
example which demonstrates this, for the case where the angle between the
spine and fan changes in time. The flow which advects the magnetic flux in the chosen example is
shown to be non-smooth at the spine and/or fan for {\it any} choice of external boundary
conditions. For typical boundary conditions (as
required by say a line-tied boundary such as the solar photosphere, or at
another null in the system) the flow will be singular. Thus non-ideal
processes are always required to facilitate such an evolution.
We presented the results of resistive MHD simulations which demonstrated that
this evolution is a generic one.

We went on to investigate the process of current sheet formation at such a 3D
null in the simulations. The null was driven by shearing motions at the spine
boundary, and a 
strong current concentration was found to result, focused at the null. The spine
and fan of the null close up on one another, from their initial orthogonal
configuration. Depending
on the strength of the driving, the current sheet may be either spread along the
fan of the null, or almost entirely contained in a spine-fan spanning sheet
(for stronger driving). 
The structure of the sheet is of exactly anti-parallel
field lines in the shear plane at the null, with the intensity of ${\bf J}$
falling off in the perpendicular direction due to the linearly increasing
field component in that direction, which is continuous across the current
sheet.  Repeating our simulations but driving at the fan
footpoints instead of the spine, a very similar evolution is observed. The
current is still inclined to spread along the fan rather than the spine for
weaker driving, consistent with relaxation simulation results
\cite{pontincraig2005}. This is natural due to the shear driving, 
and the disparity in the structures of the spine and
fan. The spine is a single line to which field lines converge, and the natural
way to form a current sheet there is to have those field lines spiral tightly
around the spine line. This field structure however is synonymous with a current directed
parallel to the spine, and is known to be induced by rotational motions,
rather than shearing ones \cite{pontingalsgaard2007}.

In addition to the current sheet at the null, indications that
non-ideal processes and reconnection should take place there are given by the
presence of a localised parallel electric field. The maxima of this parallel
electric field occur along the axis perpendicular to the shear ($z$). The
integral of $E_{\|}$ along the field line which is coincident with this axis
(by symmetry) gives the reconnection rate \cite{pontinhornig2005}, which grows
to a peak value in time before falling off after the driving is switched off.
Field lines are reconnected both across the fan and around the spine as the
null collapses.

As to the quantitative properties of the current sheet, its intensity and
dimensions are linearly dependent on the modulus of the boundary driving. 
In addition, under continual driving, we find no indications which
suggest that the sheet does not continue to grow (in intensity and length) in
time. That is, its length does not appear to be controlled by any
self-regulating mechanism. Rather, its dimensions at a given time are dependent on the boundary conditions (the degree of shearing).

Finally, with respect to a recent theorem of Eyink and Aluie \cite{eyink2006},
our results suggest 3D nulls as a possible site of turbulent
reconnection. Their theorem states that turbulent reconnection may occur at a
rate independent of the resistivity only where current sheets and vortex sheets
intersect one another. We find that strong vorticity concentrations
are in fact present in our simulations, and are localised at the null and clearly
intersect the current sheet. Moreover, in the ideal limit, our kinematic model
demonstrates that as the spine and fan collapse towards one another, the
vorticity at the fan (or spine, or both) must be singular (due to the
non-smooth transport velocity).

\section{Acknowledgements}
The authors wish to acknowledge fruitful discussions with G.~Hornig and
C-.~S.~Ng. This work was supported by the Department of Energy, Grant
No.~DE-FG02-05ER54832, by the National Science Foundation, Grant
Nos.~ATM-0422764 and ATM-0543202 and by NASA Grant No.~NNX06AC19G. 
K.~G.~was supported by
the Carlsberg Foundation in the form of a fellowship. Computations were
performed on the Zaphod Beowulf cluster which was in part funded by the Major
Research Instrumentation program of the National Science Foundation, grant
ATM-0424905.



\end{document}